\documentclass[11pt,reqno]{article}
\usepackage{amsmath,cite}
\usepackage{amsfonts}
\usepackage{amssymb}
\usepackage{amsxtra}
\usepackage{graphicx}
\usepackage{caption}
\usepackage[titletoc]{appendix}
\usepackage{ushort}
\usepackage{color}
\usepackage[hang]{footmisc}
\usepackage{hyperref}
\hypersetup{linktocpage=true}
\usepackage{float}

\makeatletter
\@addtoreset{equation}{section}
\makeatother

\def\nn{\nonumber} 
\def\ds{\documentstyle}

\newcommand{\be}{\begin{equation}}
\newcommand{\ee}{\end{equation}}
\newcommand{\bse}{\begin{subequations}}
\newcommand{\ese}{\end{subequations}}
\newcommand{\bea}{\begin{eqnarray}}
\newcommand{\eea}{\end{eqnarray}}
\newcommand{\ba}{\begin{array}}
\newcommand{\ea}{\end{array}}

\def\ft#1#2{\frac{\scriptstyle #1}{\scriptstyle #2}}
\def\fft#1#2{\frac{#1}{#2}}

\def\sst#1{{\scriptscriptstyle #1}}

\newcommand{\ie}{{\it i.e.\ }}

\newcommand{\eg}{{\it e.g.\ }}
\newcommand{\egns}{{\it e.g.}}
\def\dalemb#1#2{{\vbox{\hrule height .#2pt
        \hbox{\vrule width.#2pt height#1pt \kern#1pt
                \vrule width.#2pt}
        \hrule height.#2pt}}}

\newcommand{\hoch}[1]{$\, ^{#1}$}

\def\0{{\sst{(0)}}}
\def\1{{\sst{(1)}}}
\def\2{{\sst{(2)}}}
\def\3{{\sst{(3)}}}
\def\4{{\sst{(4)}}}
\def\5{{\sst{(5)}}}
\def\6{{\sst{(6)}}}
\def\7{{\sst{(7)}}}
\def\8{{\sst{(8)}}}
\def\9{{\sst{(9)}}}

\def\im{{{\rm i}}}

\newcommand{\mitchell}{\it George P. \& Cynthia W.
Mitchell Institute for Fundamental Physics,\\
Texas A\&M University, College Station, TX 77843-4242, USA}

\newcommand{\damtp}{\it DAMTP, Centre for Mathematical Sciences,
 Cambridge University,\\  Wilberforce Road, Cambridge CB3 OWA, UK}

\newcommand{\imperial}{\it The Blackett Laboratory, Imperial College London,\\
Prince Consort Road, London SW7 2AZ, UK }

\newcommand{\beijingnormal}{\it Center for Advanced Quantum Studies,
Department of Physics, Beijing Normal University,
Beijing 100875, China}

\newcommand{\auth}{
H. L\"u\,\footnote{mrhonglu@gmail.com}\hoch{\dagger},
A. Perkins\,\footnote{alun.perkins1@gmail.com}\hoch{\star},
C.N. Pope\,\footnote{pope@physics.tamu.edu}\hoch{\ddagger} and K.S. Stelle\,\footnote{k.stelle@imperial.ac.uk}\hoch{\star}}

\begin{document}

\pagenumbering{roman}
\setcounter{page}{0}
\thispagestyle{empty}

\begin{flushright}
Imperial/TP/17/KSS/01 \ \ MI-TH-1750
\end{flushright}

\begin{center}

\scalebox{.90}[1]{\Large {\bf Lichnerowicz Modes and Black Hole Families in Ricci Quadratic Gravity}}

\vskip 7pt

\auth

\vskip 5pt

{\hoch{\dagger}\beijingnormal}

\vskip 5pt

{\hoch{\star}\imperial}

\vskip 5pt

{\hoch{\ddagger}\mitchell}

\vskip 5pt

{\hoch{\ddagger}\damtp}

\vskip 8pt

\underline{ABSTRACT}
\end{center}

A new branch of black hole solutions occurs along with the standard Schwarzschild branch in $n$-dimensional extensions of general relativity including terms quadratic in the Ricci tensor. The standard and new branches cross at a point determined by a static negative-eigenvalue eigenfunction of the Lichnerowicz operator, analogous to the Gross-Perry-Yaffe eigenfunction for the Schwarzschild solution in standard $n=4$ dimensional general relativity. This static eigenfunction has two r\^oles: both as a perturbation away from Schwarzschild along the new black-hole branch and also as a threshold unstable mode lying at the edge of a domain of Gregory-Laflamme-type instability of the Schwarzschild solution for small-radius black holes. A thermodynamic analogy with the Gubser and Mitra conjecture on the relation between quantum thermodynamic and classical dynamical instabilities leads to a suggestion that there may be a switch of stability properties between the old and new black-hole branches for small black holes with radii below the branch crossing point.

\vfill\leftline{}\vfill
\pagebreak

\tableofcontents
\addtocontents{toc}{\protect\setcounter{tocdepth}{2}}
\newpage
\pagenumbering{arabic}
\setcounter{page}{1}
\setcounter{footnote}{0}

\section{Introduction}

  Higher derivative theories are widespread in physics, including string
theory where they arise in the low-energy expansion of the effective
action.  One can take different viewpoints as to how to treat
such higher-derivative contributions in the equations of motion.  In one viewpoint, since the
higher-derivative terms arise in a perturbative $\alpha'$ expansion,
one could take the view that only solutions where the effects of the
higher-derivative terms act as perturbations to the leading-order
form should be considered.  It may be argued that if, say,
quadratic-curvature terms were to make a significant modification to
the form of a leading-order Einstein solution, then terms of cubic and
higher order (ad infinitum) would make equally as significant further
modifications.  From such a viewpoint, focusing on a particular
set of higher-derivative corrections, while neglecting the remainder,
might seem to be unjustified.  On the other hand, one might argue that
in the vast landscape of vacua within string theory there could exist
regions where certain specific higher-order terms might play an
exceptionally dominant r\^ole. This may particularly be the case of terms quadratic in the
curvature tensor. In $n=4$ spacetime dimensions, there are just two independent such quadratic curvature invariants after excluding the Gauss-Bonnet combination $(R^{\mu\nu\rho\sigma}R_{\mu\nu\rho\sigma}-4R^{\mu\nu}R_{\mu\nu}+R^2)\sqrt{-g}$, whose $n=4$ spacetime integral is a topological invariant and so does not contribute to the perturbative structure of the theory. Inclusion of the two independent $n=4$ quadratic curvature invariants together with the Einstein-Hilbert action yields a renormalisable theory \cite{Stelle:1976gc}. The corresponding perturbative spectrum includes massless spin-two gravity together with ghost massive spin-two excitations and non-ghost massive spin-zero excitations, as well as a rich phase space of new spherically symmetric static classical solutions \cite{Stelle:1977ry, Lu:2015cqa, Lu:2015psa}. Asymptotic forms of such solutions obtained from Frobenius analysis at the origin or at horizons need to be numerically linked to linearised theory solutions at spatial infinity. Such analysis has been carried out with increased precision in \cite{Perkins:2016imn, Goldstein:2017rxn}.

   In this paper we consider the effects of a restricted set of such higher-derivative modifications to Einstein gravity in a variety of spacetime dimensions $n\ge4$. We consider the addition to the Einstein-Hilbert action of quadratic invariants in the Ricci tensor and Ricci scalar only. In other words we exclude from consideration invariants built from the full curvature such as $R^{\mu\nu\rho\sigma}R_{\mu\nu\rho\sigma}$. In $n=4$ dimensions, as noted above, this poses no limitation in generality, owing to the topological nature of the Gauss-Bonnet invariant, but in dimensions $n>4$ this restriction becomes substantive. The reason for considering only this restricted class of corrections is that Ricci-flat solutions of the leading-order theory then continue to be solutions of the corrected theory.
In particular, this means that we can continue to study certain black-hole
solutions in the corrected theory.  Much of our specific
focus will be on the case where just the quadratic invariants
$R^{\mu\nu} R_{\mu\nu}$ and $R^2$ are added, but much of our discussion will
extend, with little or no modification, to the case with higher-order
Ricci invariant terms as well.  In our opinion, it is worthwhile to study
the solution space of these theories as extensively as possible, even if
they do not necessarily capture all of the features that one might expect to
see in a fully generic higher-derivative theory.

   We shall see that negative-eigenvalue eigenfunctions of the Lichnerowicz operator play a central r\^ole
in the analysis both of static non-Schwarzschild black-hole solutions such as those found in $n=4$ dimensions in
\cite{Lu:2015cqa,Lu:2015psa} and also in the analysis of time-dependent black-hole instabilities. The existence of a bifurcation in the set
of spherically symmetric solutions to higher-derivative gravity in consequence of the $n=4$ dimensional Gross-Perry-Yaffe eigenvalue \cite{Gross:1982cv} was already presaged by Whitt \cite{Whitt:1985ki}, who also considered the black-hole stability question but did not find any classical instability. The key to finding such instabilities was found by Gregory and Laflamme \cite{Gregory:1993vy}, who considered fluctuations about an $(n+1)$ dimensional black string solution. This, in turn proves to be related to black-hole instabilities
in $n$ dimensional massive gravity theory \cite{Babichev:2013una,Brito:2013wya} and thence to black-hole instabilities in $n$ dimensional higher-derivative gravity theory \cite{Myung:2013doa}.

  In this paper, we begin in Section \ref{sec:Lichnerowiczorigin} with a study of linearised perturbations of the Ricci tensor about
initially Ricci-flat backgrounds for theories including in the action quadratic terms in the Ricci tensor and Ricci scalar. This gives rise to the study of transverse-traceless (TT) eigenfunctions of the Lichnerowicz operator generalising the Gross-Perry-Yaffe eigenfunction \cite{Gross:1982cv} in $n=4$ dimensions, which is carried out in detail in Section \ref{sec:lichopsec}. Numerical results for the analogous eigenfunctions in dimensions $4\le n\le11$ are given in Section \ref{sec:numresults}. The implications of these eigenfunctions for the existence of new spherically symmetric and static non-Schwarzschild black holes are analysed in Section \ref{sec:newblackholes} and their implications for time-dependent perturbations and stability are analysed in Section \ref{sec:timedependent}. A useful reformulation of such perturbation problems for tensor fluctuations about a Schwarzschild background in terms of a ``Schr\"odinger equation'' form was given long ago by Zerilli \cite{Zerilli:1970se}; this is applied to the present case of Ricci quadratic gravity in dimension $n$ in Section \ref{sec:Schrodinger_prob}, while application to the analysis of the threshold unstable mode is given in Section \ref{ssec:potwf}. In Section \ref{ssec:GLF}, we apply the known results on Gregory-Laflamme instabilities \cite{Gregory:1993vy} to the various $n$ dimensional cases and establish agreement between the edges of the parametric zones of instability and the Lichnerowicz negative eigenvalues given in Section \ref{sec:numresults}. The relevance of the negative Lichnerowicz eigenmodes both to the static perturbations of black holes at the Lichnerowicz crossing point and to thermodynamic instabilities leads finally in Section \ref{sec:thermodynamicimplications} to consideration of the implications of relative negative specific-heat values, as found in \cite{Lu:2015cqa}, and similarly of free energies, to suggested ranges of stability and instability for both the Schwarzschild and non-Schwarzschild black hole families.

\section{Black Holes in Ricci Quadratic Gravity}\label{sec:Lichnerowiczorigin}

  We consider the $n$-dimensional theory described by the Lagrangian\footnote{This action for Ricci quadratic gravity is given in terms of the Ricci tensor and the Ricci scalar. It is related to the $\int\sqrt{-g}(R-\alpha C_{\mu\nu\rho\sigma}C^{\mu\nu\rho\sigma} + \beta_{n=4}R^2)$ action of Refs \cite{Lu:2015cqa,Lu:2015psa} in $n=4$ dimensions by $\xi=-2\alpha$; $\beta=\beta_{n=4}+\frac23\alpha$.}
\be
{\cal L} = \sqrt{-g}\, (R + \xi\, R^{\mu\nu} R_{\mu\nu}+\beta\, R^2)\,.
\label{nlag}
\ee
Note that in $n\ge 5$ dimensions this is a specialisation compared to the most
general quadratic gravity theory, since we do not include a term
$R^{\mu\nu\rho\sigma}\, R_{\mu\nu\rho\sigma}$ in the Lagrangian.  We make this
restriction so that Ricci-flat metrics will always be solutions of the theory,
and so, of particular interest for our purposes, the Schwarzschild-Tangherlini
(hereafter just referred to as Schwarzschild) black hole will be a solution.
This will allow us to study black-hole
solutions that are perturbatively close to the Schwarzschild solution.
Without this simplifying assumption, it would be very difficult to study
black-hole solutions in the quadratic theory.

  The equations of motion following from (\ref{nlag}) are
\bea
&& R_{\mu\nu} -\ft12 R g_{\mu\nu} + 2\beta R\,(R_{\mu\nu}-\ft14 R\, g_{\mu\nu})
+(2\beta+\xi)\, (g_{\mu\nu}\, \square -\nabla_\mu\nabla_\nu)\, R\nn\\
&&
+ \xi\, \square(R_{\mu\nu}-\ft12 R \, g_{\mu\nu})
+
2\xi\, (R_{\mu\rho\nu\sigma} -\ft14 g_{\mu\nu}\, R_{\rho\sigma})\,
   R^{\rho\sigma} =0\,.\label{neom}
\eea
Taking the trace of(\ref{neom}) gives
\bea
(1-\ft12n) R + (2\beta+\xi) (n-1) \square R + \xi (1-\ft12n) \square R
+ 2(1-\ft14n)(\beta R^2 +\xi R^{\mu\nu} R_{\mu\nu}) =0\,.\label{traceeom}
\eea

  The case of $n=4$ dimensions is very special, because the
quadratic-curvature terms disappear and one is just left with the equation,
linear in $R$,
\be
2(3\beta+\xi)\, \square R -R=0\,.\label{4dtrace}
\ee
This was used in \cite{Lu:2015cqa}, following earlier work
in \cite{Nelson:2010ig} to prove that $R$ must in fact vanish for any static
spherically-symmetric black-hole solution. Essentially, one uses
a Lichnerowicz-type argument, multiplying (\ref{4dtrace}) by $R$ and
integrating over the spatial 3-volume between the horizon and infinity,
subject to the assumption of appropriate boundary conditions. This leads to
a huge simplification in the study of static black-hole solutions, allowing
the differential equations for the functions in the metric ansatz to be
reduced from 4 to 2, and thus making the use of numerical integration techniques
much easier to implement.

The findings in \cite{Lu:2015cqa} were that in
addition to the 1-parameter family of Schwarzschild black holes there exists
also a second branch of static spherically-symmetric black holes that
bifurcates from the Schwarzschild branch at a specific black-hole mass that
is determined in terms of the coefficient $\xi$ in the Lagrangian
(\ref{nlag}).

  Since in dimensions $n>4$ we shall not have the simplification of
knowing that $R=0$ for static black-hole solutions, we shall not attempt
to carry out such an extensive construction of black-hole solutions using
numerical integration methods.  What we shall still be able to do,
however, is to carry out a linearised investigation of static black-hole
solutions that are perturbatively close to the Schwarzschild solution.  These
will be the analogues of the four-dimensional black holes on the second
branch that were found in \cite{Lu:2015cqa}, in the region very close
to the bifurcation point of the two branches.  We can study such
solutions by
looking at the infinitesimal variation of the equations of motion
(\ref{neom}) around the Ricci-flat background of the Schwarzschild solution.
Making the variation, and then setting $R_{\mu\nu}=0$, we find
\bea
&&\delta R_{\mu\nu} -\ft12 g_{\mu\nu}\, \delta R +
 (2\beta+\xi)(g_{\mu\nu}\square -\nabla_\mu\nabla_\nu) \delta R
  +\xi\square\, (\delta R_{\mu\nu} -\ft12 g_{\mu\nu}\, \delta R)
       + 2\xi\, R_{\mu\rho\nu\sigma}\, \delta R^{\rho\sigma}\notag\\
        &&=0\,.
\label{veom}
\eea
Taking the trace of this equation gives
\be
(1-\ft12 n)\, \delta R +[2(n-1)\beta + \ft12 n\xi]\,\square\delta R=0\,.
\label{traceveom}
\ee
We can now again use a Lichnerowicz-type argument, in which we multiply this
by $\delta R$ and integrate over the spatial 3-volume between the black-hole
horizon and infinity.  With appropriate boundary conditions, and
for the relevant sign of the prefactor of the $\square\delta R$ term,
corresponding to a positive mass-squared $m_0^2$ for the scalar
mode, with
\be
m_0^2 = \fft{n-2}{n\xi + 4(n-1)\, \beta}\,,
\ee
we deduce that $\delta R$ must vanish.  This is weaker than the
$R=0$ result \cite{Nelson:2010ig,Lu:2015cqa} of four dimensions, but is does allow us to say that any
static black hole that is perturbatively close to Schwarzschild must still
have vanishing Ricci scalar at linearised order.  Feeding the conclusion
$\delta R=0$ back into (\ref{veom}), we then find
\be
\Big(\Delta_L  - \fft1{\xi}\Big) \, \delta R_{\mu\nu}=0\,,
    \label{LichdeltaR}
\ee
where
\be
\Delta_L\, \delta R_{\mu\nu} \equiv -\square\delta R_{\mu\nu} - 2
 R_{\mu\rho\nu\sigma}\, \delta R^{\rho\sigma}\label{Lichdef}
\ee
is the Lichnerowicz operator on a Ricci-flat background.  Note that
since $\delta R=0$
we must have $g^{\mu\nu}\, \delta R_{\mu\nu}=0$ around the Ricci-flat
background, and furthermore, from the
variation of the contracted Bianchi identity $\nabla^\mu\, R_{\mu\nu}-
 \ft12 \nabla_\nu \, R=0$ and using $\delta R=0$, we have that
$\nabla^\mu\, \delta R_{\mu\nu}=0$, so we conclude that
$\delta R_{\mu\nu}$ is a transverse traceless (TT) tensor.

Going back to (\ref{LichdeltaR}), we see that if there exists a TT tensor
eigenfunction $\psi_{\mu\nu}$ of the Lichnerowicz operator,
\be
\Delta_L\, \psi_{\mu\nu}= \lambda\, \psi_{\mu\nu} \, \label{Lichefn}
\ee
with eigenvalue given by $\lambda=1/\xi$, then we can obtain a
linearised perturbation away from the Schwarzschild solution.  In order for
the squared mass of the massive spin-2 mode $m_2^2=-1/\xi$ in the quadratic theory to be
positive (and hence non-tachyonic), we must have $\xi<0$.  Thus we will
obtain a linearised perturbation away from Schwarzschild if there exists a
negative Lichnerowicz mode, provided that $\xi$ is the inverse of the
negative eigenvalue.

As a check, we can look at the situation in four dimensions.  It is known from
the work of Gross, Perry and Yaffe \cite{Gross:1982cv} that the Schwarzschild
metric has just one normalisable mode of the Lichnerowicz operator, and that this has
$\lambda \approx -0.19\, M^{-2}$, where $M$ is the Schwarzschild mass.
Comparing with the numerical results in \cite{Lu:2015cqa}, we find that indeed
this corresponds nicely with the point at which the second branch of
solutions bifurcates from the Schwarzschild branch.

\section{Lichnerowicz Operator in a Schwarzschild Background}\label{sec:lichopsec}

  Here we give a construction of the negative-eigenvalue mode of the
Lichnerowicz operator in the background of $n$-dimensional Schwarzschild
spacetime. Consider the metric
\be
ds^2 = -h\, dt^2 +f^{-1}\, dr^2 + \rho^2\, d\Omega_{n-2}^2\,,
\ee
where $h$, $f$ and $\rho$ are functions of $r$.  Labelling the coordinates
$x^\mu=(t,r,y^i)$,
where $d\Omega_{n-2}^2 = \gamma_{ij}\, dy^i dy^j$ is the metric on the
unit $(n-2)$ sphere, the non-vanishing components of the Christoffel
connection are given by
\bea
\Gamma^0{}_{01} &=&  \fft{h'}{2h}\,,\qquad \Gamma^1{}_{11}= -\fft{f'}{2f}\,,
\qquad \Gamma^1{}_{00}= \ft12 f h'\,,\nn\\
\Gamma^1{}_{ij} &=& -\rho\rho'\,  f \gamma_{ij}\,,\qquad
\Gamma^i{}_{1j}= \fft{\rho'}{\rho}\, \delta^i_j\,,\qquad
\Gamma^i{}_{jk}= \bar \Gamma^i{}_{jk}\,,
\eea
where $\bar\Gamma^i{}_{jk}$ is the Christoffel connection for the
$(n-2)$-sphere metric $\gamma_{ij}$.  The non-vanishing components of the
Riemann tensor are given by
\bea
R_{0101}&=& \ft12 h'' -\fft{{h'}^2}{4h} + \fft{h'\, f'}{4f}\,,\qquad
R_{0i0j}= \ft12 \rho\, \rho'\,  f\, h'\, \gamma_{ij}\,,\nn\\
R_{1i1j} &=& -\Big(\rho\, \rho'' + \fft{\rho\, \rho'\, f'}{2f}\Big)\,
\gamma_{ij}\,,\qquad
R_{ijkl}=r^2\, (1-f\, {\rho'}^2)\,
(\gamma_{ik}\, \gamma_{j\ell}-\gamma_{i\ell}\,
   \gamma_{jk})\,.
\eea

   We now specialise to the case where $f=h$ and $\rho(r)=r$, and
consider a symmetric tensor
\be
\psi_{00}= h\, \psi_0\,,\qquad \psi_{11}= h^{-1}\, \psi_1 \,,\qquad
\psi_{ij}= r^2\, \gamma_{ij}\, \bar\psi\,,\label{psians}
\ee
where $\psi_0$, $\psi_1$ and $\bar\psi$ are functions of $r$ only.
We impose the tracefree condition $g^{\mu\nu}\, \psi_{\mu\nu}=0$,
which implies
\be
-\psi_0 + \psi_1 + (n-2)\, \bar\psi=0\,,\label{tracefree}
\ee
and the transversality condition $\nabla^\mu\psi_{\mu\nu}=0$, which implies
\be
\psi_1' + \fft{h'}{2h}\, (\psi_0+\psi_1) + \fft{n-2}{r}\,
(\psi_1-\bar\psi)=0\,.\label{transverse}
\ee

   The Lichnerowicz operator in a Ricci-flat background is given by
\be
\Delta_L \, \psi_{\mu\nu}= -\square\psi_{\mu\nu} -
          2 R_{\mu\rho\nu\sigma}\, \psi^{\rho\sigma}\,.
\ee
With the understanding that $\rho(r)=r$ and $f=h$, with $h$ taken to be
\be
h= 1-\Big(\fft{r_0}{r}\Big)^{n-3}\,,
\ee
giving the Schwarzschild-Tangherlini solution, we find that with $\psi_{\mu\nu}$
having the assumed form (\ref{psians}), one has
\bea
\Delta_L\psi_{00} &=& -h^2\, \psi_0'' -(h h' +\fft{n-2}{r}\, h^2)\,\psi_0'
    +\ft12 {h'}^2\,\psi_0 + (\ft12{h'}^2 - h h'')\, \psi_1\nn\\
    && -
   \fft{(n-2) h h'}{r}\, \bar\psi\,,\\
\Delta_L\psi_{11} &=& -\psi_1'' - (\fft{h'}{h} + \fft{n-2}{r})\, \psi_1'
  + (\fft{{h'}^2}{2h^2} + \fft{2(n-2)}{r^2})\, \psi_1
  +(\fft{{h'}^2}{2h^2} -\fft{h''}{h})\, \psi_0\nn\\
  &&
  +\fft{n-2}{r}\, (\fft{h'}{h} -\fft{2}{r})\, \bar\psi\,,\nn\\
\Delta_L \psi_{ij}&=& r^2\gamma_{ij}\, \Big[
 -h\bar\psi'' -(h'+\fft{n-2}{r}\, h)\bar\psi' +
   (\fft{2(n-2)}{r^2}\, h -\fft{2(n-3)}{r^2})\, \bar\psi\nn\\
   &&\qquad\quad - \fft{h'}{r}\,\psi_0
  +(\fft{h'}{r}-\fft{2h}{r^2})\, \psi_1\Big]\,.\nn
\eea
Imposing the eigenfunction condition, where for quadratic Ricci gravity one has $\lambda=1/\xi$ as in \eqref{Lichefn},
\be
\Delta_L\psi_{\mu\nu} = \lambda\, \psi_{\mu\nu}\,,\label{timedepfluctLich}
\ee
together with solving the tracefree (\ref{tracefree}) and transversality
(\ref{transverse})
conditions for $\psi_0$ and $\bar\psi$ as functions of $\psi_1$ and $\psi_1'$,
gives rise to three ordinary differential equations, coming from
the $00$, $11$ and $ij$ directions. The equation from the $11$ direction is of
second order in derivatives of $\psi_1$:
\bea
-\Big(1-\fft{r_0^{n-3}}{r^{n-3}}\Big)\, \psi_1'' &+&
  \fft{(3n^2-11n+18)\, r_0^{n+3}\, r^{n+3}-n(n-1)\, r_0^{2n}\, r^6
   -2n r_0^6\, r^{2n}}{2 r_0^6\, r^{2n+1} -(n-1)\, r_0^{n+3}\, r^{n+4}}
  \psi_1' \nn\\
&+& \fft{2n(n-3)^2\, r_0^n\, r}{2 r_0^3\, r^n - (n-1) r_0^n\, r^3}\,
  \psi_1 = \lambda\, \psi_1\,.\label{2nd}
\eea
The other two equations are of third order in derivatives of $\psi_1$, but
they are in fact implied by the second-order equation (\ref{2nd}).  Thus
it is only necessary to solve (\ref{2nd}) in order to obtain
transverse-traceless Lichnerowicz modes of the form (\ref{psians}).

   Making a trial leading-order solution of the form $\psi_1\sim (r-r_0)^c$
in the vicinity of the horizon, one finds $c=0$ or $c=-1$, and thus for
regularity we must take $c=0$.  The near-horizon expansion, which can be
used in order to set data just outside the horizon for a numerical integration
out to near infinity, then takes the form
\be
\psi_1= \sum_{m\ge 0}\, a_m\, (r-r_0)^m\,.\label{nearhor}
\ee
At large distances, when $r>>r_0$, the equation (\ref{2nd}) becomes
approximately
\be
\psi_1'' + \fft{n}{r}\, \psi_1' + \lambda\, \psi_1\approx 0\,,
\ee
and thus for a negative eigenvalue we have
$\psi_1\sim e^{\pm \sqrt{-\lambda}\,r }$ at large $r$.  The numerical
solution of the equation can therefore be performed by the shooting method,
seeking a value of the eigenvalue $\lambda$ for which the exponentially
diverging solution $e^{+ \sqrt{-\lambda}\,r }$ is absent.  As in
four dimensions, there appears in a general dimension to be just
one negative-eigenvalue normalisable mode.

\section{Numerical Results}\label{sec:numresults}

  In this section we universally set $r_0=1$, so that the Schwarzschild
black hole has unit horizon radius.  We constructed a near-horizon Taylor
expansion of the form (\ref{nearhor}), up to $(r-1)^8$ order, and used this
to set initial data at a radius $r_i=1+1/100$ just outside the horizon, for
numerical integration out to large $r$.  If the eigenvalue parameter
$\lambda$ is taken to be positive, one finds as expected oscillatory
solutions for all $\lambda>0$. It appears to be the case for each dimension
$n\ge4$ that there is just one negative value of $\lambda$ for which
one can have a normalisable eigenfunction (i.e. no exponentially-growing term
$\sim e^{\sqrt{-\lambda}\, r}$).  For dimensions $4\le n \le 11$ we find
\bea
n=4:&& \lambda\approx -0.7677\,,\nn\\
n=5:&& \lambda \approx -1.610\,,\nn\\
n=6:&& \lambda\approx -2.499\,,\nn\\
n=7:&& \lambda\approx -3.417\,,\nn\\
n=8:&& \lambda \approx -4.356\,,\nn\\
n=9:&& \lambda\approx -5.309\,,\nn\\
n=10:&& \lambda\approx -6.272\,,\nn\\
n=11:&&\lambda\approx -7.242\,.\label{staticevs}
\eea
The $n=4$ result accords with the eigenvalue reported by Gross, Perry and Yaffe \cite{Gross:1982cv}.
(They took $M=1$ and hence $r_0=2$; the eigenvalues scale like $1/r_0^2$ if
one removes the $r_0=1$ specialisation we took.)  The $n\ge5$
results are broadly
in accordance with the higher-dimensional results reported from previous
work in Eqn (34) of Ref.\ \cite{Myung:2013cna}.

\section{New Black Holes}\label{sec:newblackholes}

\subsection{Transverse and traceless gauge}

The existence of a TT Lichnerowicz mode with negative eigenvalue
$\lambda$ in the Schwarzschild background implies the existence of a
new branch of static,
spherically-symmetric black-hole
solutions  that bifurcates from the Schwarzschild black-hole branch at the point
where
$\xi=1/\lambda$.  To see this, we note that for the transverse and
traceless perturbation of the Schwarzschild metric,
$g_{\mu\nu}\rightarrow g_{\mu\nu} + h_{\mu\nu}$ with
$g^{\mu\nu} h_{\mu\nu}=0=\nabla^\mu h_{\mu\nu}$, we have $\delta R_{\mu\nu}=
\ft12 \Delta_L h_{\mu\nu}$, implying that the equations of motion become
\be
(\Delta_L - \lambda)\Delta_L h_{\mu\nu}=0\,.
\ee
(The analogous equation in AdS black hole background was obtained in \cite{Liu:2011kf}.)
The general solution of this fourth-order equation is a linear superposition
of solutions of
$\Delta_L h_{\mu\nu}=0$ and solutions of
$(\Delta_L - \lambda)h_{\mu\nu}=0$.

    For perturbations that
maintain the spherical symmetry, solutions with $\Delta_L h_{\mu\nu}=0$
lead again to a Schwarzschild black hole. In fact the Lichnerowicz TT
zero modes can easily be solved for using the results presented
in Section \ref{sec:lichopsec}.  The general solution of (\ref{2nd}) for
the function $\psi_1$, in the case $\lambda=0$, is given by
\be
\psi_1 = \fft{c_1\, [2 r^{n-3} - (n-1)(n-2)]}{(1-r^{n-3})} +
  \fft{c_2\, [(n-1)-2(n-2)\, r^{n-3}]}{r^{n-1}\, (1-r^{n-3})}\,,
\ee
where $c_1$ and $c_2$ are arbitrary constants.  (We have again set $r_0=1$
here for convenience.)  From this, it follows, in particular, that
\be
\psi_{ij}= \, -2 (c_1\, r^2 + c_2\, r^{3-n})\, \gamma_{ij}\,,
\ee
and so the perturbation $g_{\mu\nu}\rightarrow g_{\mu\nu} + \psi_{\mu\nu}$
can be re-expressed in terms of a canonical radial variable $\rho$ for which
$ds^2= \cdots +\rho^2\, d\Omega_{n-2}^2$ by means of the redefinition
\be
r = \rho + c_1\, \rho + c_2\, \rho^{2-n}\,,\label{diffeo}
\ee
where $c_1$ and $c_2$ are now understood to be infinitesimal. A
straightforward calculation of the rest of the perturbed metric shows that
the parameter $c_2$ is trivial (corresponding merely to a diffeomorphism that
is absorbed by the redefinition (\ref{diffeo})), and that the parameter
$c_1$ corresponds to a mass perturbation under which the original
Schwarzschild metric function $1-\rho^{3-n}$ is transformed according to
\be
1-\fft1{\rho^{n-3}}\longrightarrow 1-\fft{1 + c_1\, (n-1)(n-3)}{\rho^{n-3}}
\,.
\ee

  On the other hand, perturbations with $(\Delta_L-\lambda) h_{\mu\nu}=0$, with
$\xi=1/\lambda$, describe a branch of
new black-hole solutions.  To see this concretely, we write the general
spherically symmetric ansatz for static black holes
\be
ds^2 = - h dt^2 + f^{-1}\, dr^2 + \rho^2 d\Omega_{n-2}^2\,,
\ee
where $(h,f,\rho)$ are functions just of $r$.  It is then straightforward to see
that for a transverse and traceless perturbation of the form (\ref{psians}),
the new black-hole solution is given by
\be
h= \bar f (1 - \epsilon \psi_0)\,,\qquad f= \bar f (1 - \epsilon \psi_1)\,,
\qquad
\rho=r(1 + \ft12 \epsilon \bar \psi)\,,\label{pertnoScbh}
\ee
with $\bar f=1-r_0^{n-3}/r^{n-3}$ for the unperturbed background
Schwarzschild black hole.  In this perturbation, the horizon remains at
$r=r_0$, but the area of the horizon is changed, being given by
\bea
{\cal A}=\ft14 r_0^{n-2}\,\Omega \rightarrow \ft14 r_0^{n-2}
(1 +\ft12\epsilon\bar \psi)^{n-2}\,\Omega\,,
\eea
where $\Omega$ is the spherical volume factor of the $(n-2)$-dimensional space $d\Omega_{n-2}^2$.

\subsection{Fixed horizon area}

In an alternative description, we may consider a gauge choice in which the area of
the horizon is fixed.  This can be done by requiring
\be
\rho=r\,,\qquad
h=\bar f (1 + \epsilon\, \tilde h)\,,\qquad
f=\bar f (1 + \epsilon\, \tilde f)\,.
\ee
Equations (\ref{Lichefn}) taken together with $\delta R=0$ yield
\bea
&&2r^2 \bar f\tilde h'' - r \big((n-5)\bar f-3(n-3)\big)\tilde h' +
r\big( (n-3)\bar f + n-3\big) \tilde f' +2(n-2)(n-3)\tilde f= 0\,,\cr
&&-4r^2\big(r\bar f \tilde h' + (n-3) \tilde f\big) +\xi\Big(2r^2\bar f
\big((n-1)\bar f-n+3\big) \tilde f''\cr
 &&\qquad-r \big((n-1)\bar f^2 - 2(n-3)(3n-5)\bar f +
5(n-3)^2\big)\tilde f'\cr
&&\qquad + r(n-3)(\bar f-1) \big((n+1)\bar f-n+3\big)\tilde h'\cr
&&\qquad +2(n-3)\big((n^2-5n+2)\bar f - (n-2)(n-3)\big) \tilde f\Big)=0\,.
\eea
Note that we have fixed the radial coordinate gauge choice by imposing
$\rho(r)=r$.  This means that the metric perturbation is neither
transverse nor traceless in this gauge.
In order for the perturbative solution to be regular on the horizon, we consider
the Taylor expansion
\be
\tilde f=1 + f_1 (r-r_0) + f_2 (r-r_0)^2 + \cdots\,,\qquad
\tilde h=h_0 + h_1 (r-r_0) + h_2 (r-r_0)^2 + \cdots\,.
\ee
with
\be
f_1=\fft{1}{r_0}\Big(1-\ft12n - \fft{3r_0^2}{4(n-3)\xi}\Big)\,,\qquad
h_1=\fft{1}{r_0}\Big(1-\ft12n + \fft{r_0^2}{4(n-3)\xi}\Big)\,,\qquad\hbox{\it etc.}
\ee
For each given $r_0$, there is indeed a value of $\xi$ for which a
solution exists.  For example, with $r_0=1$, we find such a value in
$n=6$ dimensions to be
\be
\xi\sim -0.4002\,,\qquad \fft{1}{\xi}=-2.499\,.
\ee
The large-$r$ behaviour of the function $\tilde f$ is
\bea
\tilde f &=& \fft{c_1}{r^3} + \fft{c_2(r^2 + 3\mu_0^2 r + 3\mu_0^2)e^{-r/\mu_0}}{r^3} +\cdots\,,\cr
\tilde h &=& \fft{c_1}{r^3} + \fft{4c_2\mu_0(r+\mu_0)e^{-r/\mu_0}}{r^3} +\cdots\,.
\eea
We find that $h_0=1.401$ and $c_1=3.602$.  The coefficient $c_2$ is too
unstable to be determined accurately owing to the exponential factor.

\subsection{Non-Schwarzschild black holes below and above the branch crossing}\label{sec:smallnSbh}

One feature of the black-hole solution space that becomes manifest from the Lichnerowicz eigenfunction analysis of the static spherically symmetric perturbations away from the Schwarzschild solution is that the non-Schwarzschild solutions \eqref{pertnoScbh} exist both for smaller as well as larger black holes than the solution sitting precisely at the crossing of the two branches. In other words, the perturbation parameter $\epsilon$ controlling location along the non-Schwarzschild branch in \eqref{pertnoScbh} can be negative as well as positive.

\begin{figure}[h]
\centering
\includegraphics[scale=.70]{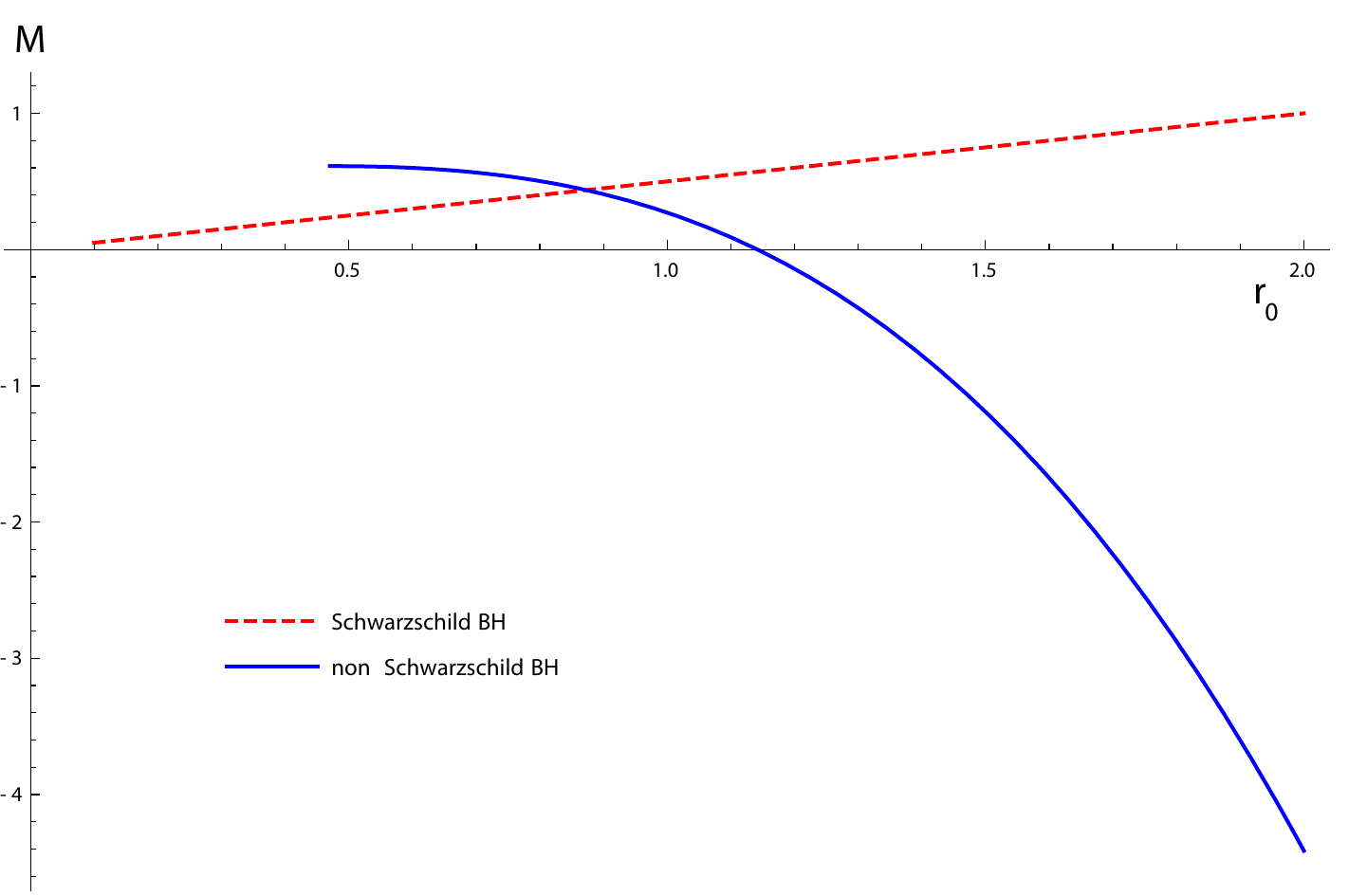}
\caption{Non-Schwarzschild black-hole solutions crossing the Schwarzschild branch.}
\label{fig:BH_trajectories}
\end{figure}

The possibility of such sub-crossing as well as super-crossing non-Schwarzschild black holes was not highlighted in the earlier numerical analysis of Ref.\  \cite{Lu:2015cqa}, but once their existence is understood from the Lichnerowicz perturbation analysis, it is possible to find them numerically \cite{Perkins:2016imn, Goldstein:2017rxn}. Using the same sort of numerical analysis as in Refs \cite{Lu:2015cqa, Lu:2015psa}, one obtains in $n=4$ spacetime dimensions the branch structure shown in Figure \ref{fig:BH_trajectories}.

For black holes significantly smaller than that at the branch crossing, numerical calculation becomes increasingly unstable as one moves to smaller $r_0$, and it is not yet clear what happens for very small $r_0$ although it seems that the non-Schwarzschild branch asymptotes to a limiting mass value.

\subsection{The first law of thermodynamics}

Having obtained linearised numerical solutions of new black holes that can
be viewed as small perturbations from the Schwarzschild black hole, we
can establish the first law of black-hole thermodynamics by use of
the Wald formalism.  The Wald formalism for spherically-symmetric black
holes in quadratically extended gravity in general dimensions was derived in Ref.\ \cite{Fan:2014ala}.
The entropy can be derived from the Wald entropy formula,\footnote{The Wald entropy is given \cite{Wald:1993nt,Iyer:1994ys} by $S=-\fft{1}{8} \int_+ \sqrt{h}\, d^{n-2}x\, \epsilon^{ab}\epsilon^{cd}\, \fft{\partial L}{\partial R^{abcd}}$ where the integration is over a cross-section of the horizon with metric $h_{\mu\nu}$ and binormal $\epsilon^{ab}$.} giving
\bea
S &=& \ft14{\cal A} \Big(1 - (n-2) (4\beta + \xi) \fft{f'(r_0)\rho'(r_0)}{\rho(r_0)} +
\fft{2(n-2)(n-3)\beta}{\rho(r_0)^2}\cr
&& + \ft14 (2\beta + \xi)
\big(f''(r_0) + \fft{3f'(r_0) h''(r_0)}{h'(r_0)}\big) \Big)\,,
\eea
where the derivative is with respect to $r$ and $r=r_0$ is the location of
the horizon.  For the Schwarzschild black hole with horizon radius $r_0$,
the mass, temperature and entropy are
\be
M=\fft{(n-2) r_0^{n-3}\,\Omega}{16\pi}\,,\qquad
T=\fft{n-3}{4\pi r_0}\,,\qquad S=\ft14 r_0^{n-2}\,\Omega\,,
\ee
If the Schwarzschild solution is perturbed by making an infinitesimal
variation of the parameter $r_0$, the corresponding changes in
$M$ and $S$ obey the first law $dM=TdS$.

We now consider the perturbative description of the new
branch of black-hole solutions, firstly in the gauge
where the metric perturbation is transverse and traceless.
In this case, since the normalisable perturbative mode falls off
exponentially at large $r$, the mass is unchanged by the perturbation, and
hence $\delta M=0$ as we move along the trajectory of perturbative
deformations.  The near-horizon expansion of the metric functions are
\bea
h &=& h_0(r) (1 - \epsilon \psi_0) \sim (n-3)(1+\epsilon)(r-1)\cr
&& -\Big(\ft12(n-2)(n-3) +\ft12\epsilon \big((n-3)(3n-2) +\ft3{\xi}\big)\Big)(r-1)^2+ \cdots\,,\cr
f &=& h_0(r) (1-\epsilon \psi_1 r)\sim (n-3)(1-\epsilon)(r-1)\cr
&& -\Big( \ft12(n-2)(n-3) -\ft12\epsilon \big((n-3)(3n-2) +\ft3{\xi}\big)\Big)(r-1)^2 + \cdots\,,\cr
\rho &=& r - \fft{\epsilon}{n-2} + \Big(1 + \fft{(n-1)(n-3)\xi + 1}{(n-2)(n-3)\xi}\Big)(r-1) + \cdots
\eea
Substituting these into the entropy formula, we find
\be
S\sim \ft14\Omega \Big(r - \fft{\epsilon}{n-2}\Big)^{n-2} (1 + \epsilon) \sim \ft14{\cal A}_0 + {\cal O}(\epsilon^2)\,.
\ee
Thus $\delta S=0 + {\cal O}(\epsilon^2)$, implying that $S$ is also
unchanged at linear order in $\epsilon$, as we move along the trajectory of
deformations.  This means that in the transverse-traceless
gauge, the first law $\delta M=T\delta S$ is trivially verified as one
moves along the new branch of black holes (at the linearised level),
since both $M$ and $S$ are unchanged as the perturbation is
turned on.

  In the alternative description in the gauge where $\rho(r)=r$
(fixed horizon area),
the temperature and entropy of the new black hole are
\bea
T&=& \ft{\sqrt{h'(r_0)f'(r_0)}}{4\pi} = \fft{n-3}{4\pi r_0}\Big(1 + \fft{\epsilon (1 + h_0)\xi}{2r_0^2}\Big)\,,\cr
S &=& \ft14 {\cal A} \Big[1 - (n-2) (4\beta + \xi) \fft{f'(r_0)}{r_0} +
\fft{2(n-2)(n-3)\beta}{r_0^2}\cr
&&\qquad\qquad - \ft14 (2\beta +\xi) \big(f''(r_0) +
\fft{3f'(r_0)h''(r_0)}{h'(r_0)}\big)\Big]\cr
&=& \ft14 r_0^{n-2} \Omega \Big(1 + \fft{\epsilon (n-2)(n-3)\xi}{r_0^2}\Big)\,.
\eea
Thus we see that the entropy of the perturbed solution differs from that of the
original
solution by
\be
\delta S=\ft14\epsilon\, (n-2)(n-3)\, r_0^{n-4}\xi \Omega\,,
\ee
This implies from the first law that we must have
\be
\delta M=\fft{\epsilon\,\Omega}{16\pi} (n-2)(n-3)^2 \xi r_0^{n-5}\,.\label{deltam}
\ee
We do not have a universal proof, but we can establish the relation
by numerical analysis in a case-by-case basis.  For the $n=6$
numerical solution described in the previous section, we find
\be
M=\fft{\Omega}{4\pi} (1 - 3.60\, \epsilon)\,,\qquad
\delta M=-3.60\, \epsilon\,  \fft{\Omega}{4\pi}\,.
\ee
This is precisely the $\delta M$ from the first law (\ref{deltam}),
with $n=6$, $\xi=-0.4002$ and $r_0=1$.

\section{Time-Dependent Perturbations and Stability}\label{sec:timedependent}

   So far, we have discussed spherically-symmetric TT perturbations that
are time independent.  In particular, one finds there is one such
negative-eigenvalue mode of the Lichnerowicz operator in the Schwarzschild
background.  The approximate values for the negative eigenvalue in dimensions
$4\le n\le 11$ are given in Eqn (\ref{staticevs}) for the case where the
Schwarzschild radius $r_0$ is taken to be unity.  If we denote the
negative eigenvalues $\lambda$ for such time-independent solutions in $n$ dimensions by $\bar\kappa_n$ when $r_0=1$,
then for general values of $r_0$ the negative eigenvalues will be
\be
\kappa_n = \fft{\bar\kappa_n}{r_0^2}\,.\label{kappa_n}
\ee

  We now consider the stability properties of the black hole solutions. Once again, the Lichnerowicz negative eigenvalue will play a key r\^ole, this time in delimiting the edge of a domain of instability for the Schwarzschild black-hole family. We begin with a formulation setting up a perturbative approach.

\subsection{Time-dependent Fluctuation Problem in Schr\"odinger Form}\label{sec:Schrodinger_prob}

   In four dimensions, a way to recast into a Schr\"odinger form for a single field the Lichnerowicz problem of time-dependent tensor fluctuation modes around a Schwarzschild background was given by Zerilli \cite{Zerilli:1970se}. This formulation has been particularly useful in analysing instabilities in massive gravity theories \cite{Brito:2013wya}. Here, we give the analogous results for perturbations about a Schwarzschild background in arbitrary dimensions for Ricci quadratic gravity.

   Including time dependence, which we take to be of the form $e^{-i\omega t}=e^{\nu t}$,
the spherically-symmetric ansatz for TT modes takes the form
\be
\psi_{00}= h\, \psi_0(r)\, e^{\nu t}\,,\quad
\psi_{01}= \chi(r)\, e^{\nu t}\,,\quad \psi_{11}= h^{-1}\, \psi_1(r)\,
   e^{\nu t}\,,\quad \psi_{ij}= r^2\, \bar\psi(r)\, \gamma_{ij}\, e^{\nu t}\,.\label{timedepmetric}
\ee
The traceless and transverse conditions imply three non-trivial equations,
which can be solved for $\psi_0$, $\bar\psi$ and $\chi'$.  The
Lichnerowicz eigenvalue equation
\be
\Delta_L\, \psi_{\mu\nu}=\lambda \psi_{\mu\nu}\label{timedepLich}
\ee
implies two two-derivative
equations (the $(01)$ and $(11)$ components), and feeding in the
expressions for $\psi_0$, $\bar\psi$ and $\chi'$ from the TT conditions,
we can solve these for $\chi$ itself, and obtain a second-order
equation purely for $\psi_1$.  The remaining two non-trivial
equations from the Lichnerowicz eigenvalue equation are of third order
in derivatives of $\psi_1$, and these can be seen to be consequences of the
second-order equation for $\psi_1$.

   To cast the equation into a Schr\"odinger form, we follow steps
analogous to those that can be used in four dimensions to reproduce
the Zerilli form.  Thus, we introduce a new variable $\phi(r)$, defined
by
\be
\phi(r) = \nu^{-1}\, u(r)\, \chi(r) + v(r)\, \bar\psi(r)\,,
\ee
where $u$ and $v$ are functions of $r$ that will be determined shortly.
Crucially, we assume that $u(r)$ and $v(r)$ do not depend on $\nu$ (although
they may depend on the Lichnerowicz eigenvalues $\lambda$, as well as
$r$).\footnote{These requirements are motivated by the specific example
of the Zerilli construction in four dimensions, where these properties
hold.}
We solve for $u$ and $v$ by making the following requirements.  Firstly,
we require that in terms of the ``tortoise'' coordinate $r_*$, ranging from the horizon as $r_*\to-\infty$ to spatial infinity as $r_*\to+\infty$ and defined by
\be
r_* = \int^r \fft{dr'}{h(r')}\,,
\ee
the function $\phi$ should satisfy an equation of the form
\be
\fft{d^2\phi}{dr_*^2} = W(r)\, \phi\,,\label{schrod1}
\ee
where we make use of the second-order equation satisfied by $\psi_1$ that
was determined previously.  The requirement that the left-hand side of
(\ref{schrod1}) should be expressible purely as a function of $r$ times
$\phi$ (\ie that there should be no term with a first derivative of $\phi$)
implies one equation with derivatives of $u$ and $v$ (up to second order).
The second requirement we make is that $W(r)$ should be of the form
\be
W(r) = \nu^2 + V(r)\,,
\ee
where $V(r)$ does not depend on $\nu$.  The consequent property that
$dV/d\nu=0$ gives an equation (up to first derivatives only) involving
$u$ and $v$.  This equation is exactly integrable, leading to the
relation
\be
v(r) = \fft{r\, u(r)}{h(r)}\,.
\ee
Feeding this back into the previous second-order equation that followed from
the first of the two above requirements results in an integrable
equation for $u$, whose solution is
\be
u(r) = \fft{h(r)\, r^{\ft32 (n-2)}}{\ft12(n-2)(n-3) -\lambda r^{n-1}}\,.
\ee
Note that here, and below, we are taking the Schwarzschild radius to be
$r_0=1$, so
\be
h(r) = 1-\left(\fft{r_0}{r}\right)^{n-3}=1 -\fft{1}{r^{n-3}}\,.
\ee

    With $u$ and $v$ now determined, we have arrived at the Schr\"odinger
form of the Lichnerowicz eigenvalue equation
\be
-\fft{d^2\phi}{dr_*^2} +\Big[ \nu^2 + V(r)\Big]\, \phi=0\,,\label{Schrodeigenvalueeqn}
\ee
where the Zerilli-type potential $V(r)$ is given by
\be
V(r)= -\fft{h(r)}{r^{n-1}\, [\ft12(n-2)(n-3) -\lambda\, r^{n-1}]^2}\,
       Y(r)\,,\label{Zerillipot}
\ee
where
\bea
Y(r) &=& -\ft1{16} (n-2)^3\, (n-3)^2\, [(n-2) + (n-4) r^{n-3}] \nn\\
&&+
  \ft14 (n-2)(n-3)\lambda\, r^{n-1}\, [2n^2-5n+6 -3n(n-2) r^{n-3}] \nn\\
&&
+\ft14 (n+2)\, \lambda^2\, r^{2(n-1)}\, [3(n-2) - n r^{n-3}]  +
  \lambda^3\, r^{3(n-1)}\,.\label{Yfn}
\eea
\newpage

\subsection{Potentials and Wavefunctions}\label{ssec:potwf}

   The Schr\"odinger form \eqref{Schrodeigenvalueeqn} for the equation governing the
metric perturbations is a convenient starting point for an analysis of
the stability of the static spherically-symmetric black-hole
solutions.  The equation cannot be solved analytically, and so it is
necessary to resort to approximation techniques or to numerical analysis.
It is however straightforward to obtain solutions numerically, and
this seems to be the simplest way to proceed.

\begin{figure}[H]
\centering
\includegraphics[scale=.7]{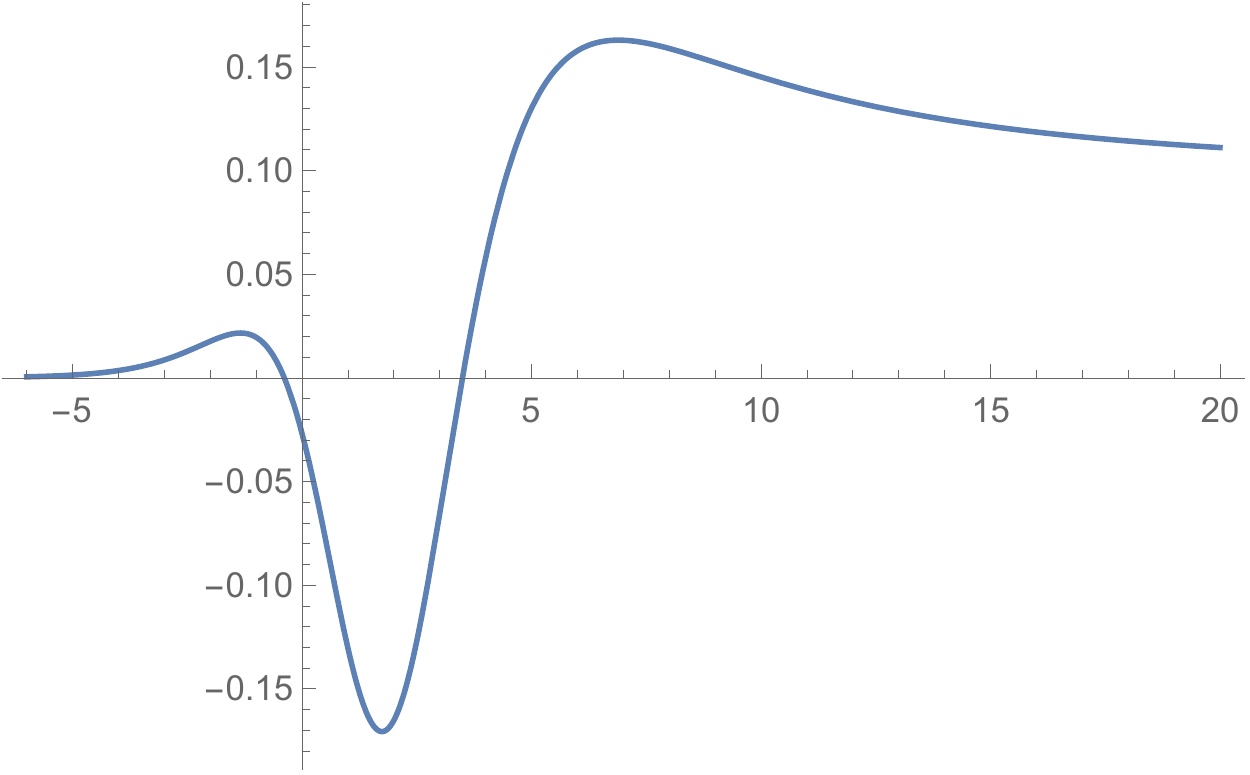}
\caption{\it Zerilli-type potential $V(r_*)$ in $n=4$ dimensions for the Schr\"odinger problem.}
\label{fig:genZerillipotential}
\end{figure}

First, however, we need to determine the relevant boundary conditions for the Schr\"odinger wavefunction $\phi$ solving
\eqref{Schrodeigenvalueeqn}. In general, the Zerilli-type potential $V$ \eqref{Zerillipot} has a form when expressed in terms of the $r_*$ tortoise coordinates like that shown in Figure \ref{fig:genZerillipotential}.
Key features of this potential are its asymptotic limits to zero as $r_*\to-\infty$ and to a nonzero constant as $r_*\to+\infty$.  The analysis of black-hole quasinormal modes for nonzero frequencies given in References \cite{Schutz:1985zz,Iyer:1986np} prescribes purely outgoing waves in the left and the right asymptotic regions, \ie asymptotic coordinate dependences of the forms  $e^{-i\omega(r_*+t)}$ as $r_*\to-\infty$ and $e^{i\omega(r_*-t)}$ as $r_*\to+\infty$.\footnote{For generic values of $\lambda$ below a certain limiting value, Figure \ref{fig:genZerillipotential} shows a mild ``hump'' maximum before settling down to its asymptotic $r_*\to+\infty$ constant value. The WKB analysis of quasinormal modes given in Refs \cite{Schutz:1985zz,Iyer:1986np} matches the top of a similar ``hump'' of a Zerilli-type potential for perturbations of the Kerr solution to a parabola and then uses parabolic cylinder functions to do the WKB matching and so derive the imaginary parts of quasinormal-mode frequencies. Our main concern here, however, is to understand the boundary of the domain of instability for the Schwarzschild solution, \ie the limiting $\lambda$ value of the solution at which the instability disappears. As it happens, the potential ``hump'' in the present problem diminishes in this limit and a standard WKB analysis becomes unreliable.} The limiting wavefunction at the boundary of the domain of instability has $\nu=i\omega=0$ and so must tend to $r_*$ independent constants as $r_*\to\pm\infty$. Since $V(r_*)\to0$ as $r_*\to-\infty$, any constant asymptotic value of $\phi$ will satisfy \eqref{Schrodeigenvalueeqn} in that region, but for $V(r_*)\to\text{const}\ne0$ as $r_*\to+\infty$, one must have $\phi(r_*)\to0$ as $r_*\to+\infty$ in order to satisfy \eqref{Schrodeigenvalueeqn}. Accordingly, the tortoise-coordinate outgoing-wave boundary conditions are equivalent to regularity conditions as $r_*\to\pm\infty$. These regular boundary conditions can be satisfied only for specific values of $\lambda$ in the \eqref{Zerillipot} potential $V(r_*)$, depending on the dimension $n$.

\begin{figure}[ht!]
\begin{center}
\includegraphics[scale=.4]{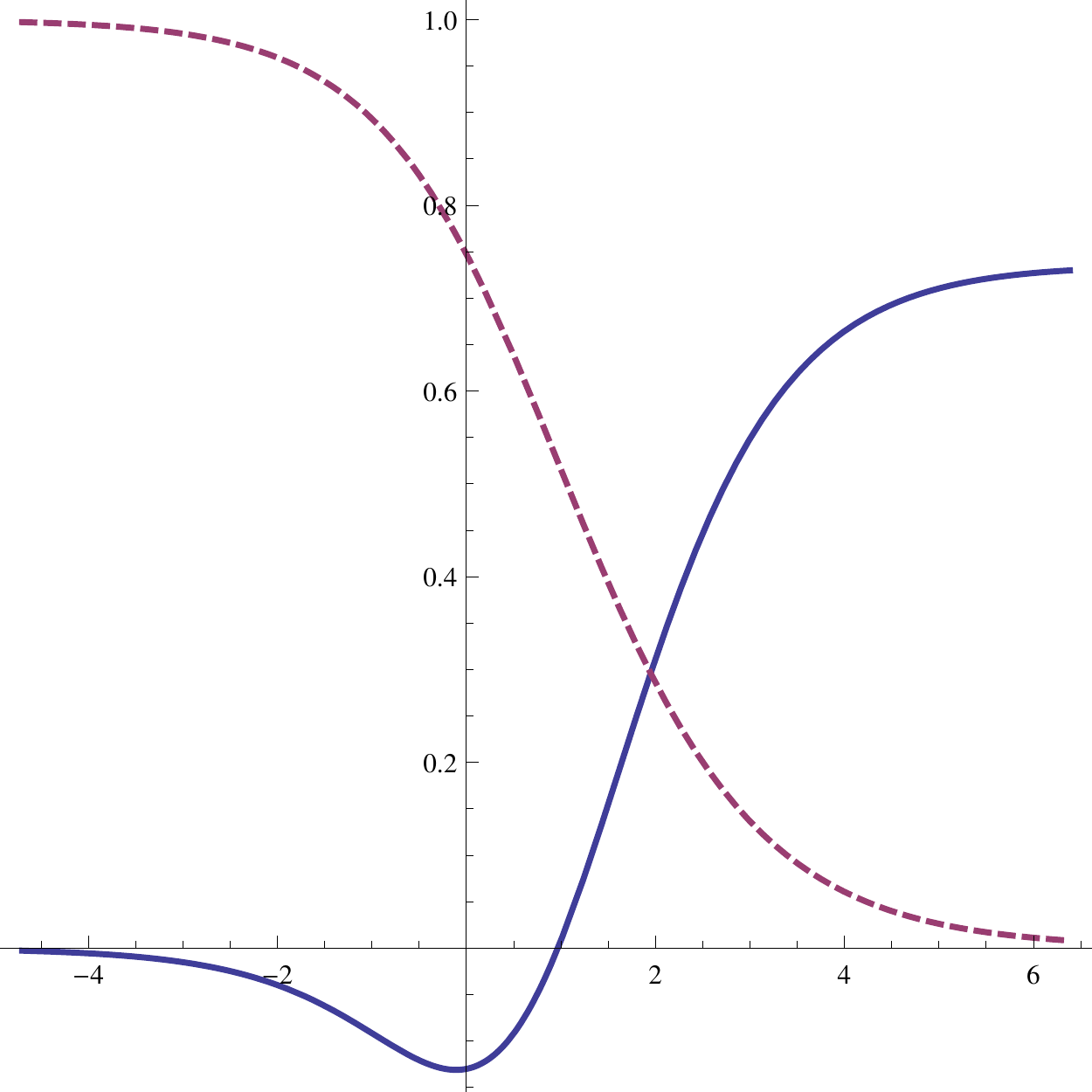}\ \
\end{center}
\caption{\it The Zerilli-type potential $V(r_*)$ (solid line), and the
wavefunction $\phi(r_*)$ (dashed line) in $n=4$ dimensions for the Lichnerowicz negative eigenvalue
$\lambda\approx -0.7677$ that gives rise to a regular solution.}\label{d4fig}
\end{figure}

  As an illustration, consider the regular $\nu=0$ solution in $n=4$ dimensions at the edge of the domain of instability.  By means of a simple shooting-method numerical study of the regular solution of equation \eqref{Schrodeigenvalueeqn}, we find that this is achieved when the constant $\lambda$ takes the specific value $\lambda=\bar\kappa_4\approx-0.7677$, in full agreement with the results given earlier in Section \ref{sec:numresults}.  In Figure \ref{d4fig} we display the potential $V$, plotted as a function of the
redefined radial coordinate $r_*$, together with a plot of the
corresponding wavefunction $\phi$.   The potentials and wavefunctions
are qualitatively similar in all dimensions.

\subsubsection{Square well approximation}

\begin{figure}[ht!]
\begin{center}
\includegraphics[width=200pt]{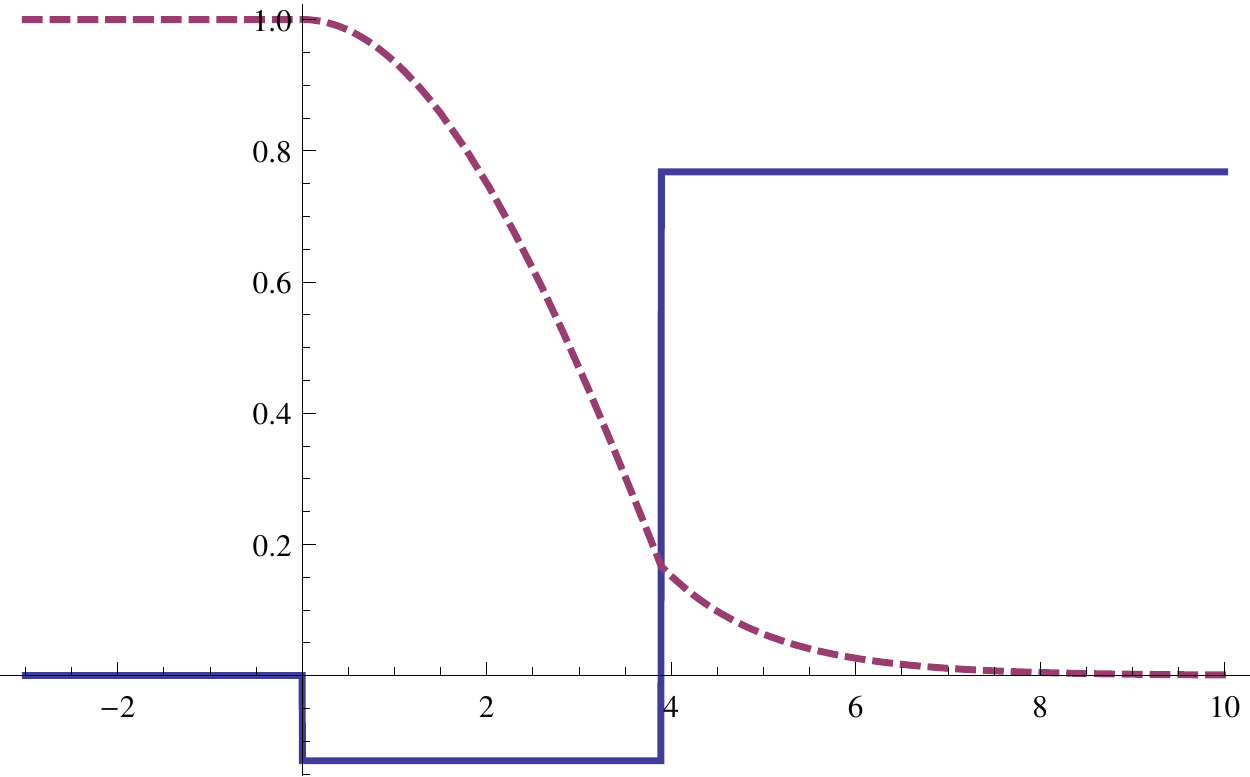}\ \
\end{center}
\caption{\it The rectangular well potential (solid line), and the wavefunction (dashed line), for the regular solution in $n=4$ dimensions.}\label{rectwellfig}
\end{figure}

  Unlike the Zerilli-type potential for the case of perturbations in
standard Einstein gravity, \eg for the Kerr solution, it can be seen that the potential shown in Figure
\ref{d4fig} does not readily lend itself to a straightforward WKB approximation.  Another approximate
approach is to consider a simple idealisation of the potential in which it is represented by
a rectangular well, as depicted in Figure \ref{rectwellfig}.  The approximation of the true potential \eqref{Zerillipot} by a rectangular well provides an instructive qualitative picture of the nature
of the regular solutions to the Schr\"odinger-type equation \eqref{Schrodeigenvalueeqn}. Let the rectangular-potential analogue of the Schr\"odinger wavefunction $\phi$ be denoted $\varphi$.

The specific
details of the well in this figure have been chosen for the four-dimensional
case, but the same general idea can be applied in all dimensions.  Thus
in general one can define a rectangular potential
\be
V(x)= \begin{cases} 0\ \hbox{for}\  x<0\,,  \\
              V_-\ \hbox{for}\ 0\le x\le \ell\,, \\
              V_+\ \hbox{for}\ x>\ell\,,
      \end{cases}
\ee
where $V_-<0$ and $V_+>0$.  The coordinate $x$ here corresponds to
the radial coordinate $r_*$ in the Schr\"odinger formulation
in \eqref{Schrodeigenvalueeqn}, up to an unimportant additive constant.  The
values of $V_-$ and $V_+$ are chosen to match roughly the corresponding
values of the minimum and maximum of the actual potential given by
\eqref{Zerillipot} and \eqref{Yfn}.

As we discuss below,
for generic values of the  quantities $V_-$, $V_+$ and $\ell$ that
characterise the rectangular potential, the solution for
 the exact Schr\"odinger wavefunction will necessarily be divergent either
as $x$ goes to $-\infty$ (corresponding to the horizon) or
as $x$ goes to $+\infty$ (corresponding to $r\rightarrow\infty$) or both.
Only by fine tuning the parameters of the potential can one obtain a
solution where the wavefunction is regular everywhere.  This is analogous to
the way in which the parameter $\lambda$ in the Zerilli-type potential
$V(r)$ in the actual Schr\"odinger problem $-d^2\phi/dr_*^2 + V\phi=0$
is determined by the requirement that $\phi$ should be regular both on the
horizon and at infinity.

   Concretely, if we define
\be
V_-= - k^2\,,\qquad V_+= -\lambda =\mu^2\,,
\ee
then the general solutions $\varphi_1$, $\varphi_2$ and $\varphi_3$ in the three
regions $x<0$, $0\le x\le \ell$ and $x>\ell$ respectively are:
\bea
\varphi_1 &=& a_1 + b_1\, x\,,\nn\\
\varphi_2 &=& a_2\, \cos k x + b_2\, \sin k x\,,\nn\\
\varphi_3 &=& a_3 \, e^{-\mu x} + b_3 \, e^{\mu x}\,.
\eea
  In order for the wave function to be finite as $x\rightarrow -\infty$ we must have
$b_1=0$.  Without loss of generality, we shall then choose the normalisation
$a_1=1$.  To match the functions and first derivatives at $x=0$ we then
need $a_2=1$ and $b_2=0$.
 For the function $\varphi_3$ to be regular as
$x\rightarrow\infty$, we must have $b_3=0$.  Finally, to match the
functions and first derivatives at $x=\ell$ we require
\be
\cos k\ell = a_3\, e^{-\mu\ell}\,,\qquad k \sin k\ell = \mu\,  a_3\,
e^{-\mu\ell}\,.\label{xell}
\ee
Thus we have
\be
\varphi_1=1\,,\qquad \varphi_2= \cos k x\,,\qquad \varphi_3= a_3\, e^{-\mu x}\,,
\ee
where $\mu$, $k$ and $\ell$ must satisfy
\be
\tan k\ell = \fft{\mu}{k}\,,\label{taneq}
\ee
and $a_3$ is then determined from (\ref{xell}).  We can see graphically
how the solutions to
(\ref{taneq}) arise.
Writing (\ref{taneq}) in terms of $V_+$, $V_-$ and $\ell$, we have
\be
V_+ = (-V_-)\, \tan (\sqrt{-V_-}\, \ell)\,.\label{taneq2}
\ee
This is the condition on the parameters of the rectangular potential that
must be satisfied in order to obtain an everywhere-regular wavefunction.

   We can now compare the results from this rectangular-well approximation
with the actual results from the numerical solutions.  Using
$V_+=-\lambda$, and reading off $V_-$ from the plot in Figure \ref{d4fig},
we can
then use (\ref{taneq2}) to calculate $\ell$, the width of the negative
part of the rectangular potential, finding
\be
V_+= 0.7677\,, \quad V_- = -0.13\,, \quad \implies \ell\approx 3.89\,.
\ee
We see that this width for the negative part of the rectangular potential
does indeed match reasonably well with the estimate one can make from
the plot in figure \ref{d4fig}, and that the wavefunction in the rectangular
approximation reasonably matches that in the true numerical solution.

\subsection{Gregory-Laflamme instabilities}\label{ssec:GLF}

The instability of the Schwarzschild solution in $n$-dimensional quadratic Ricci gravity as discussed in the previous subsection is directly related to a familiar instability of the black string in $n+1$ dimensions, as found by Gregory and Laflamme \cite{Gregory:1993vy}.

The result of the Gregory-Laflamme analysis is that for each dimension $n$ there exists a continuous
family of spherically-symmetric time-dependent
TT eigenfunctions with negative eigenvalues $\lambda$ lying in the
range
\be
\kappa_n < \lambda <0\,,\label{unstable}
\ee
where $\kappa_n$ is as given in \eqref{kappa_n} and for which the time dependence has a
real exponential form $e^{\nu\, t}$, thus corresponding to
instabilities growing exponentially in time.  That is to say, the
static negative eigenmode discussed in Sections \ref{sec:lichopsec} and  \ref{sec:numresults} represents also the
limiting, \ie most negative, value of a continuous range of less negative eigenvalues for TT modes with real exponential time dependence. For this reason, the static mode may be called a ``threshold unstable mode''.
%
%

  Related discussions can be given for quadratic Ricci gravity or for pure massive gravity or for which the same form of equation applies \cite{Myung:2013doa}. The black-string metric in $(n+1)$ dimensions is given
by simply adding a $dz^2$ term to the black-hole metric in $n$ dimensions:
\be
d\hat s^2 = ds^2 + dz^2\,.\label{glfmetric}
\ee
The relevant S-wave in $n$ dimensions fluctuations will be of the form
\be
\hat \psi_{\mu\nu} = e^{\im\, \mu\, z}\, \psi_{\mu\nu}\,,
\ee
where the TT field $\psi_{\mu\nu}$ depends just on $r$ and $t$ of the the $n$-dimensional
spacetime, with the $t$ dependence assumed to be of the form \eqref{timedepmetric}. For the $n\times n$ part of the metric $\psi_{\mu\nu}$ we then have the Lichnerowicz condition
\be
(\Delta_L + \mu^2)\, \psi_{\mu\nu}=0\,,\label{redmetricLich}
\ee
which clearly shows the identity between the $(n+1)$ dimensional black-string time-dependent fluctuation problem and the quadratic Ricci gravity time-dependent fluctuation problem \eqref{timedepfluctLich} with the identification $\mu^2=m_2^2=-\lambda=-\frac1\xi$.  Accordingly, the Gregory-Laflamme black-string instabilities of Ref.\ \cite{Gregory:1993vy} directly determine the instabilities of the quadratic Ricci gravity theories. They also directly determine the instabilities of black holes in massive gravity theory with spin-two mass $m_2$. Moreover, the eigenvalue interval \eqref{unstable} for instabilities of the $(n+1)$ dimensional black string problem translates directly into the corresponding interval for instabilities of the Schwarzschild solution in quadric Ricci gravity.

Fix a value $\xi$ for the
coefficient of the $R^{\mu\nu}\, R_{\mu\nu}$ term in the Lagrangian \eqref{nlag}.  We then
know that there will be an associated critical value $r_c$
of the Schwarzschild radius given by
\be
\fft1{\xi} = \fft{\bar\kappa}{r_c^2}\,,\label{rcdef}
\ee
corresponding to the radius at which the bifurcation of static black-hole branches occurs and the
second branch of non-Schwarzschild black holes crosses the Schwarzschild
branch. Now consider a Schwarzschild black hole of radius $r_0$ in
this theory (with $\xi$ related to $r_c$ as in (\ref{rcdef})).
Since the TT fluctuations satisfy $\Delta_L \,\psi_{\mu\nu}= (1/\xi)\,
\psi_{\mu\nu}$, it follows that the criterion (\ref{unstable}) for
having an instability implies
\be
\fft{\bar\kappa_n}{r_0^2} < \fft{\bar\kappa_n}{r_c^2}\,.
\ee
Since $\bar\kappa_n$ is negative, this means that $1/r_0^2 > 1/r_c^2$, and hence
the Schwarzschild black hole will be unstable if its radius $r_0$ is
such that
\be
r_0 < r_c\,.
\ee
Thus Schwarzschild black holes whose radius is less than the radius
at the bifurcation point become unstable, while those whose radius exceeds
that at the bifurcation point are stable.

 The black-string analysis also gives directly information about black-string instabilities in massive gravity theories. Consider a massive
gravity theory in $(n+1)$ spacetime dimensions for which the TT fluctuations satisfy the Lichnerowicz condition
\be
(\hat\Delta_L + m_2^2 )\, \hat\psi_{MN}=0\,,
\ee
where $m_2$ is the mass of the spin-2 field in $(n+1)$ dimensions.
The relevant fluctuations will again be of the form
\be
\hat \psi_{\mu\nu} = e^{\im\, \mu\, z}\, \psi_{\mu\nu}\,,
\ee
where the $n\times n$ TT field $\psi_{\mu\nu}$ depends just on $r$ and $t$ of the the $n$-dimensional
spacetime.  In $n$ dimensions we then have the condition
\be
(\Delta_L + \mu^2 + m_2^2)\, \psi_{\mu\nu}=0\,.
\ee

   If we take $ds^2$ to be the metric of an n-dimensional Schwarzschild black hole of
radius $r_0$, there will accordingly be exponentially-growing time-dependent modes
if
\be
\fft{\bar \kappa_n}{r_0^2} < -\mu^2 -m_2^2\,.
\ee
In view of the fact that $\bar\kappa_n$ is negative, this means that
the criterion for instability is $(-\bar\kappa_n)/r_0^2> \mu^2+m_2^2$,
and so conversely, the criterion for {\it stability} can be written as
\be
r_0^2 \ge \fft{(-\bar\kappa_n)}{(\mu^2+m_2^2)}\,.
\ee
Thus we will have complete stability (for all possible $\mu$) if
\be
r_0^2 \ge \fft{(-\bar\kappa_n)}{m_2^2}\,.\label{stability}
\ee

If we consider ordinary massless gravity with $m_2=0$, the criterion
(\ref{stability}) can never be satisfied, regardless of the size of the
black hole.  This gives rise to the usual Gregory-Laflamme instability for
black strings in $(n+1)$ dimensional Einstein gravity.  However, if we consider instead black strings
in $(n+1)$ dimensional massive gravity theory, then we see that provided the $n$ dimensional Schwarzschild
black hole has a sufficiently large radius such that (\ref{stability}) is
satisfied, then the black string is protected from a Gregory-Laflamme
instability.

\section{Thermodynamic Implications for Stability}
\label{sec:thermodynamicimplications}

An important issue for a variety of black-hole systems has been the relation between thermodynamic instability and spacetime dynamical instability \cite{Gubser:2000ec,Gubser:2000mm,Reall:2001ag,Hollands:2012sf,Wald:2014bia}. This is clearly germane to the context of the present paper as one can see from the r\^ole of the Gross-Perry-Yaffe Lichnerowicz negative eigenvalue in setting the boundary of the Schwarzschild dynamical S-wave instability, whereas it originally arose in the analysis of thermodynamic instability in Euclideanised quantum gravity.

One message of these studies is that spacetime dynamical instability requires thermodynamic instability, although the converse is not always the case. For systems with extensive quantities including ADM mass $M$ together with other extensive quantities $X_i$, such as angular momentum $J_i$, ADM spatial momentum $P_i$ or charges $Q_i$, dynamical stability is equivalent \cite{Wald:2014bia} to thermodynamic stability on the subspace of perturbations satisfying
\be
\delta M = \delta X_i =0\,.\label{waldstabilityreq}
\ee
In pure Einstein theory, although the static Schwarzschild solution is thermodynamically unstable at the quantum level owing to the existence of the Gross-Perry-Yaffe negative Lichnerowicz eigenvalue, this does not imply classical dynamical instability because this spherically symmetric Schwarzschild solution does not possess any other extensive conserved quantities $X_i$, and the thermodynamically unstable perturbation is not consistent with the sole $\delta M=0$ requirement. The situation changes for stationary Kerr black holes where the presence of nonzero angular momentum $J_i$ allows for thermodynamically and dynamically unstable perturbations satisfying \eqref{waldstabilityreq} to exist, leading, \egns, to the phenomenon of superradiance \cite{Brito:2013wya,Green:2015kur}.

In systems such as black branes, which in their static configurations are spatially homogeneous, instabilities can occur via perturbations which locally disturb this homogeneity while maintaining overall conservation of extensive quantities such as total mass or angular momentum \cite{Hollands:2012sf}. Accordingly in systems such as the $n=5$ black string of Eq.\ \eqref{glfmetric}, instabilities can occur that involve ``bunching'' of the mass distribution in the $z$ direction while maintaining spherical symmetry in the $n=4$ dimensional subspace. This ``bunching'' mechanism does not apply to solutions without such homogeneous structure, but a criterion for stability against perturbations subject to \eqref{waldstabilityreq} can be given generally in terms of the positivity requirement of a ``canonical energy'' quantity as given in Ref.\ \cite{Hollands:2012sf}.

The analysis of relations between thermodynamic and dynamical instabilities in Einstein theory, including with matter coupling, has not yet been extended to the cases of massive gravity or of higher-derivative gravity. However the relevance of the negative Lichnerowicz eigenvalue both to the existence of the new family of non-Schwarzschild black holes and to the edge of the zone of Schwarzschild instability as discussed in Section \ref{sec:timedependent} suggests that such a thermodynamic-dynamic instability correspondence may still apply in these cases. A full study of dynamical stability could require a numerical quasinormal mode analysis about both Schwarzschild and non-Schwarzschild families of solutions, which in the non-Schwarzschild case is itself only known numerically. However, one can nonetheless gain a hint of the relative stability properties of the two black hole families from thermodynamic considerations.


Two features of the thermodynamic-dynamic instability correspondence in Einstein theory lead to an analogous suggestion in quadratic curvature gravity.

\begin{itemize}
\item
One is the existence of negative specific heats for solutions from the thermodynamic point of view. The key relation here is between the absolute values of the specific heats for Schwarzschild and non-Schwarzschild black holes in regions where the classical stability properties of the Schwarzschild solution are known from the Gregory-Laflamme analysis.
\item The other is the necessary presence of threshold unstable modes as one crosses from a parametric zone of instability to a zone of stability \cite{Reall:2001ag}. Such a necessarily time-independent mode also constitutes the linearised perturbation describing the branching of the non-Schwarzschild black hole family as one moves from $r_0 < r_c$ instability to $r_0 > r_c$ stability (with $r_c$ as given in \eqref{rcdef}) along the Schwarzschild family trajectory as shown in Figure \ref{fig:BH_trajectories}.
\end{itemize}

Conversely, as one moves along the non-Schwarzschild family trajectory of Figure \ref{fig:BH_trajectories}, crossing a boundary between an unstable zone and a stable zone would similarly require a threshold unstable mode for this family from the thermodynamic point of view. From the dynamical point of view, this would correspond simply to a perturbation away from the crossing point onto the Schwarzschild family.

The question then arises as to the direction in which the change from stability to instability might take place. An indication of an answer to this question may be taken from the relative magnitudes of the specific heats.

It was pointed out already in \cite{Lu:2015cqa} that in Ricci quadratic gravity, both the Schwarzschild and non-Schwarzschild black holes have negative specific heats, \ie negative values of $C=dM/dT$. In that paper, discussion was given only of the portion of the non-Schwarzschild solution branch with masses M less than that of the Lichnerowicz crossing point, and it was pointed out that for such solutions, the non-Schwarzschild black holes have specific heats $C$ that are more negative than those of the corresponding Schwarzschild solution at the same temperature $T$. This suggests that the non-Schwarzschild black holes lying to the right of the branch crossing point in Figure \ref{fig:BH_trajectories} may be unstable there, given the stability of the Schwarzschild black holes in that range.

\begin{figure}[H]
\centering
\includegraphics[scale=.70]{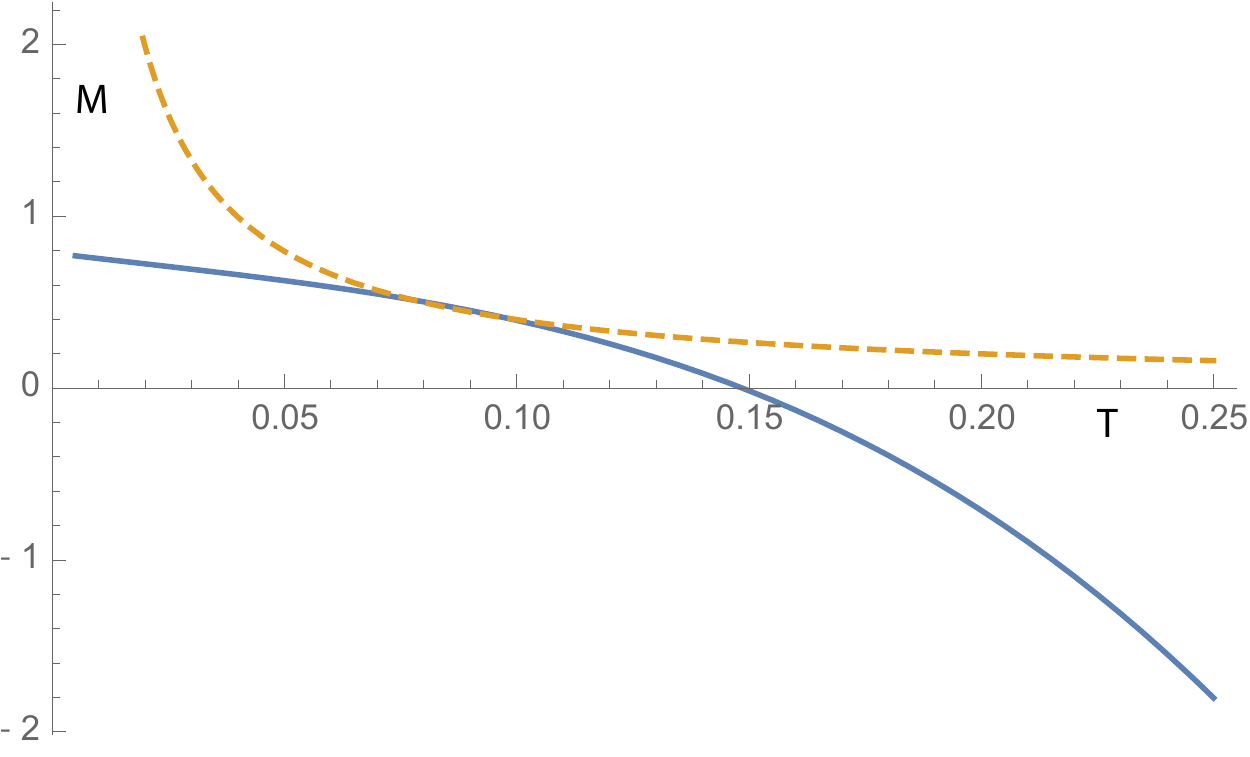}
\caption{\it Mass $M$ versus temperature $T$ relations for Schwarzschild (dashed line) and non-Schwarzschild (solid line) black holes.}
\label{fig:MT_rels}
\end{figure}

\begin{figure}[H]
\centering
\includegraphics[scale=.70]{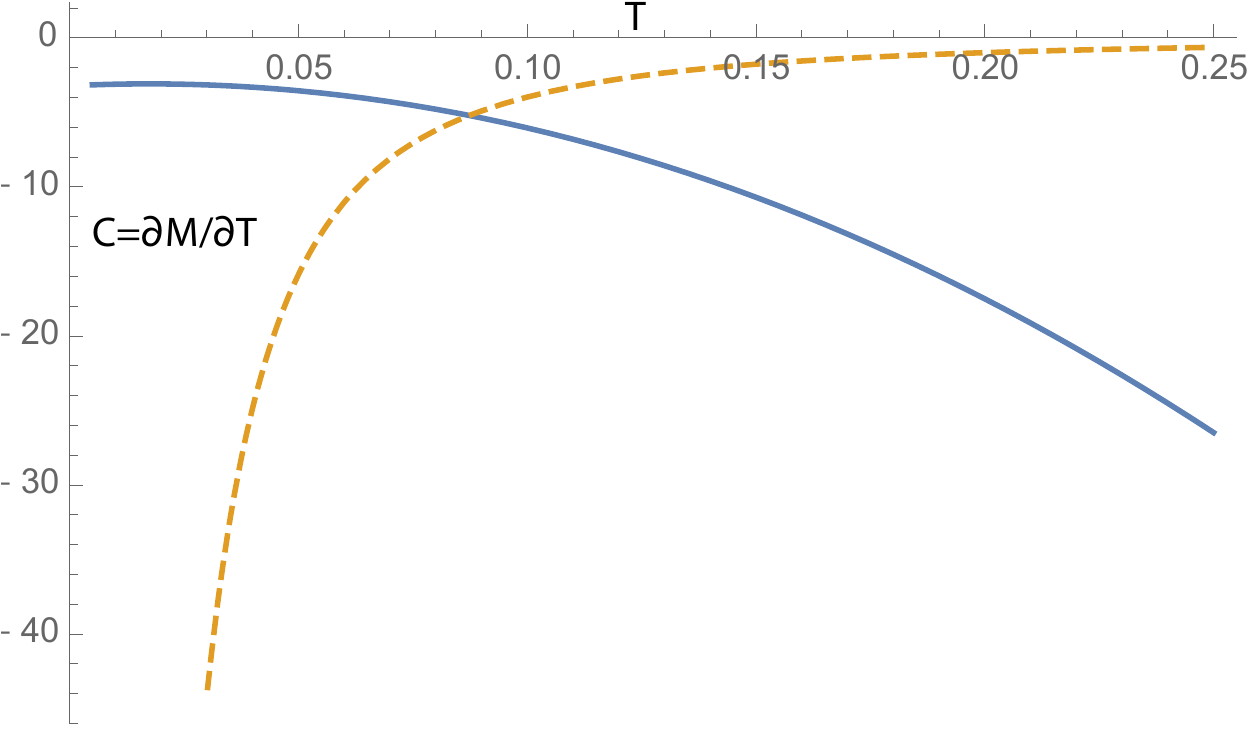}
\caption{\it Specific heat $C$ versus temperature $T$ relations for Schwarzschild (dashed line) and non-Schwarzschild (solid line) black-holes families.}
\label{fig:Specific_heats}
\end{figure}

What happens, then, in the range of solutions lying to the left of the Figure \ref{fig:BH_trajectories} crossing point? We may gain an idea of the relative susceptibility of the two branches to instabilities by considering a plot of the corresponding specific heats. In Reference \cite{Lu:2015cqa}, for the $n=4$ dimensional case, the relations between Wald entropy $S$ \cite{Wald:1993nt,Iyer:1994ys} and mass $M$ and temperature $T$ were given for the non-Schwarzschild black holes as
\bea
M_{\hbox{\scriptsize NSch}}&\approx& 0.168 + 0.131 \, S - 0.00749 \, S^2 - 0.000139\, S^3
    +\cdots\,,\nn\\
 T_{\hbox{\scriptsize NSch}} &\approx& 0.131 - 0.0151 \, S - 0.000428 \, S^2 + \cdots\,.\label{MTS}
\eea
Eliminating $S$ yields the mass versus temperature relation for non-Schwarzschild black holes
\be
M_{\hbox{\scriptsize NSch}} = 0.785221  -3.22043 \, T + 7.49256 \, T^2 - 144.089 \, T^3 + \cdots\ ,\label{MTNSch}
\ee
while for Schwarzschild black holes, one has the classic relation
\be
M_{\hbox{\scriptsize Sch}}=\frac1{8\pi T}\ ;\label{MTSch}
\ee
a plot of the results is shown in Figure \ref{fig:MT_rels},
\noindent while for the corresponding specific heats $C=dM/dT$ one has the results shown in Figure \ref{fig:Specific_heats}.

From Figure \ref{fig:Specific_heats}, one observes that for higher temperatures, which for Schwarzschild black holes correspond to small masses $M$ and small radii $r_0$ and which for non-Schwarzschild black holes correspond to smaller (and eventually negative) masses but larger radii, the non-Schwarzschild black holes have a more negative specific heat than the Schwarzschild black holes. Accordingly, since Schwarzschild black holes are known to be subject to classical Gregory-Laflamme instabilities in this portion of their family trajectory, the suggestion is that the non-Schwarzschild black holes are more unstable in this ``hot'' portion of their family trajectory than the Schwarzschild black holes.\footnote{In this discussion, we adopt the helpful ``hot'' and ``cold'' terminology of Ref.\ \cite{Goldstein:2017rxn}.}

Conversely, for lower temperatures, which for Schwarzschild black holes correspond to large masses $M$ and large radii $r_0$ and for non-Schwarzschild black holes correspond to larger (now all positive) masses but smaller radii, the non-Schwarzschild black holes have a less negative specific heat than the Schwarzschild black holes. Accordingly, since Schwarzschild black holes are known to be classically immune to Gregory-Laflamme instabilities in this portion of their family trajectory, the suggestion is that the non-Schwarzschild black holes are more stable than the in this ``cold'' portion of their family trajectory than the Schwarzschild black holes.

\begin{figure}[H]
\centering
\includegraphics[scale=.70]{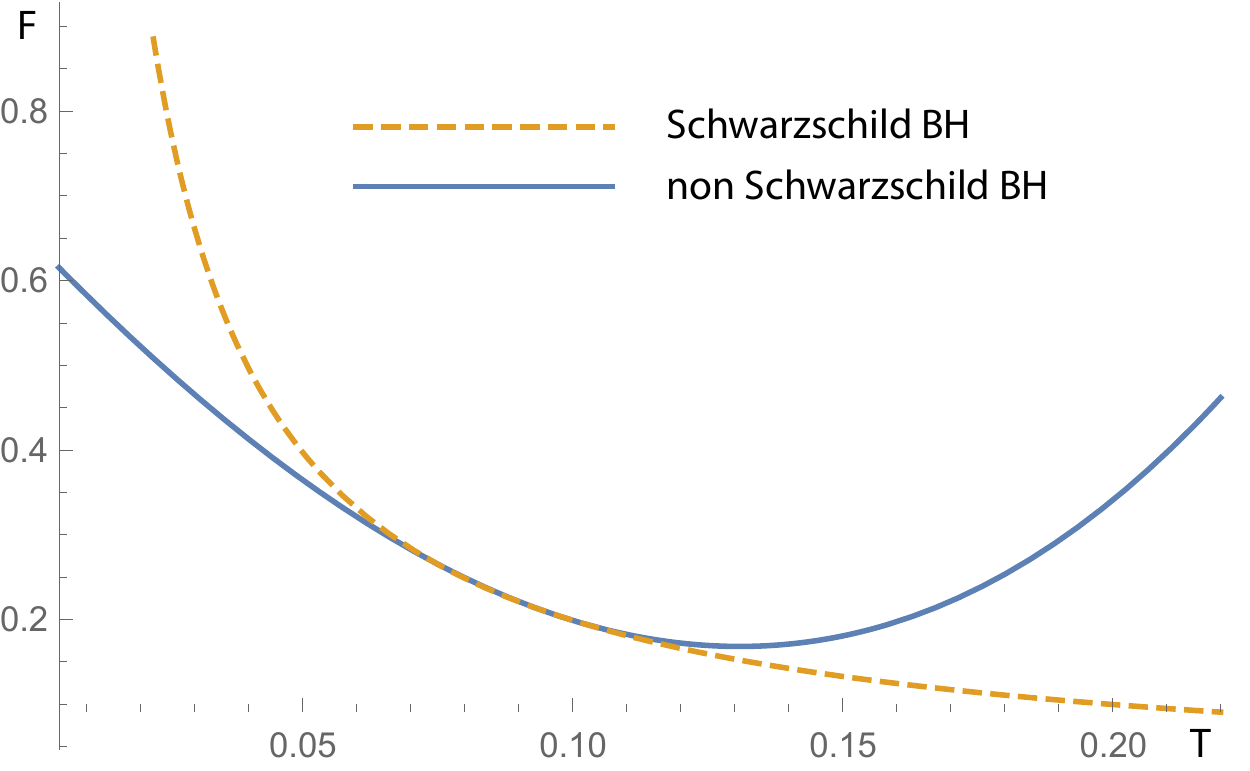}
\caption{\it Free energy $F=M-TS$ versus temperature $T$ relations for Schwarzschild (dashed line) and non-Schwarzschild (solid line) black holes families.}
\label{fig:Free-energy}
\end{figure}

The same inferences may be drawn by considering a graph of the free energy $F=M-TS$ versus temperature $T$ relations of the two black-hole families, as shown in Figure \ref{fig:Free-energy}, where again one sees a crossing of the two family curves.\footnote{A graph of the free-energy versus temperature relations for Schwarzschild and ``hot'' non-Schwarzschild black holes was given in Ref.\ \cite{Lu:2015cqa}. Extension of the non-Schwarzschild solution family to the ``cold'' segment at radii below the crossing point was given in Ref.\ \cite{Perkins:2016imn} without focussing on thermodynamic aspects. A graph analogous to Figure \ref{fig:Free-energy} showing the crossing of the free-energy curves was recently given in Ref.\ \cite{Goldstein:2017rxn}.}
At temperatures above the crossing point of the two black-hole family curves, where the Schwarzschild solution is known to have Gregory-Laflamme instabilities, the free energy of the non-Schwarzschild solutions is higher, indicating a greater susceptibility to thermodynamic instability. Conversely, at temperatures below the crossing point, where the Schwarzschild solution is classically stable, the free energy of the non-Schwarzschild solutions is lower, indicating greater thermodynamic stability.

\begin{figure}[H]
\centering
\includegraphics[scale=.70]{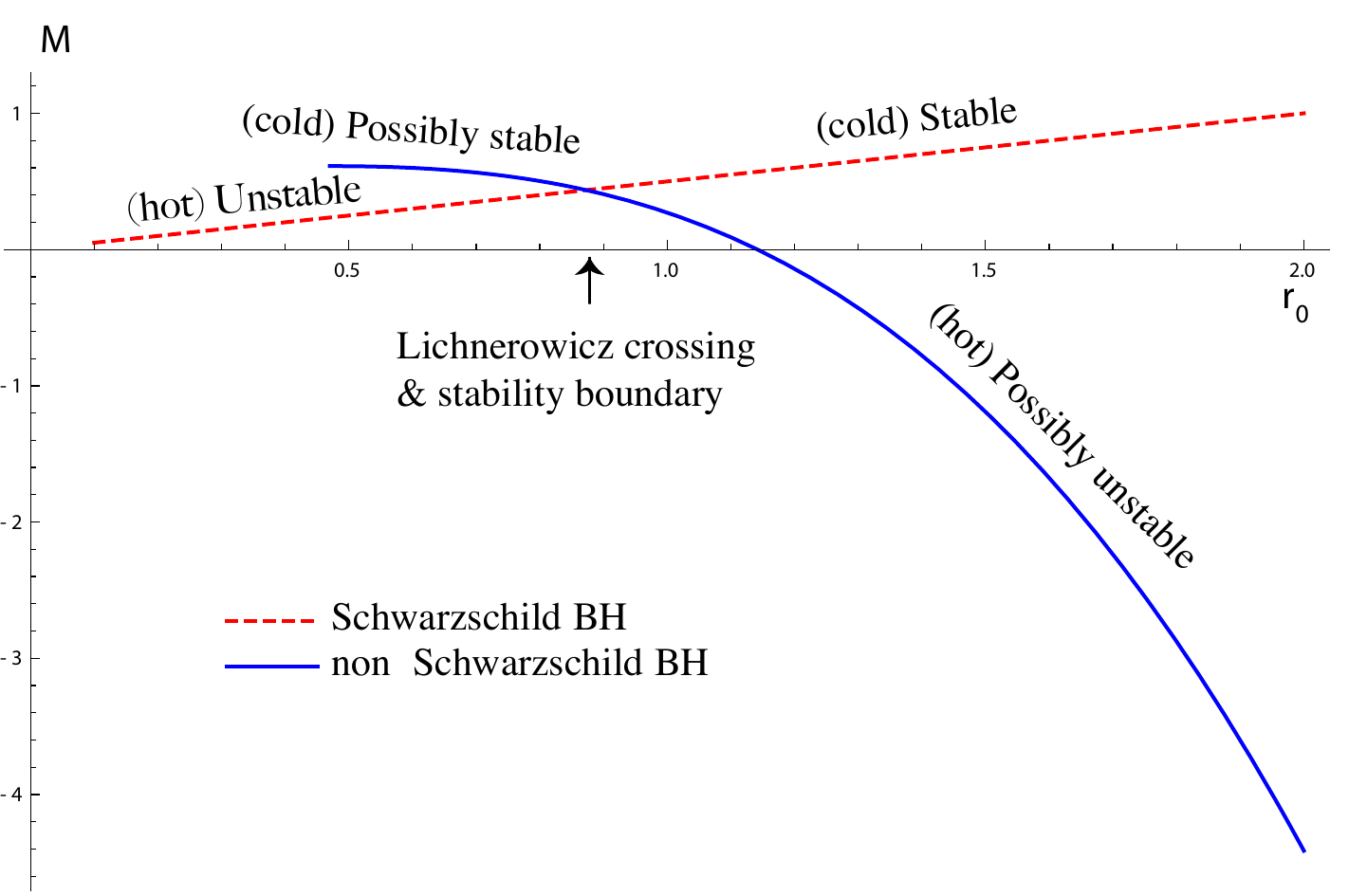}
\caption{\it Classical stability ranges for Schwarzschild (dashed line) and suggested classical stability ranges for non-Schwarzschild (solid line) black holes.}
\label{fig:Stability_ranges}
\end{figure}

Thus, a coherent suggestion emerges for the phase structure of dynamical stability and instability ranges of the Schwarzschild and non-Schwarzschild black hole families, as shown in Figure \ref{fig:Stability_ranges}. This suggestion arises from two interrelated observations. The first is the existence of threshold unstable modes in linear perturbations away from the Schwarzschild and the non-Schwarzschild black hole families at the Lichnerowicz crossing point, as shown in the analysis of Sections \ref{sec:Lichnerowiczorigin}, \ref{sec:lichopsec} and  \ref{sec:numresults}. The second is the pattern of relative susceptibilities to thermodynamic instability as revealed by study of the specific heats or the free energies of the solution families.

\section{Conclusion}
\label{sec:conclusion}

We have seen how the static negative-eigenvalue Lichnerowicz eigenfunction plays an interrelated set of r\^oles related to black-hole solutions of Ricci quadratic gravity in dimensions $n\ge4$.  Writing the fluctuation equation \eqref{LichdeltaR} in terms of the metric when expanded about a Ricci-flat background, one has
\be
\Delta_L(\Delta_L+m_2^2)\delta g_{\mu\nu}=0\,,\label{eq:metricfluct}
\ee
so one has both $\Delta_L\delta g_{\mu\nu}=0$ Lichnerowicz zero-mode solutions and $(\Delta_L+m_2^2)\delta g_{\mu\nu}=0$ Lichnerowicz negative-mode solutions. These provide leading-order perturbative branching between the family of conventional Schwarzschild black holes and the non-Schwarzschild black holes found in Refs \cite{Lu:2015cqa,Lu:2015psa}. The negative mode also forms a threshold unstable mode boundary between known stable ``cold'' and unstable ``hot'' Schwarzschild phases, the unstable phase being subject to S-wave Gregory-Laflamme classical dynamical instabilities.

The suggestion made in Section \ref{sec:thermodynamicimplications} from analysis of the relative values of specific heats or free energies for the Schwarzschild and non-Schwarzschild solution branches is that the $\Delta_L\delta g_{\mu\nu}=0$ mode may play a similar dual r\^ole for the non-Schwarzschild branch: it obviously provides a leading-order branching off the non-Schwarzschild branch onto the conventional Schwarzschild branch, and conjecturally could also form a threshold unstable mode boundary between stable ``cold'' and unstable ``hot'' phases of the non-Schwarzschild branch. 

What one learns from the thermodynamic analysis is comparative: at ``hot'' temperatures where the Schwarzschild black holes are classically dynamically unstable and have high free energies, implying thermodynamic instability, the non-Schwarzschild black holes have even higher free energies, suggesting a similar classically dynamically unstable phase for them as well. Conversely, at ``cold'' temperatures where the Schwarzschild black holes are classically dynamically stable and have low free energies, implying greater thermodynamic stability, the non-Schwarzschild black holes have even lower free energies, suggesting a similar classically dynamically stable phase for them.

Confirmation of this suggested stability phase structure could be made by a direct analysis of the quasinormal modes, or by further development of the relation between quantum thermodynamic stability and classical dynamical stability for higher-derivative gravity theories, along the lines of the canonical energy analysis of References \cite{Hollands:2012sf,Wald:2014bia}.

The present paper has focussed exclusively on Ricci quadratic gravity models \eqref{nlag} in various dimensions $n$, which all admit classic Schwarzschild-Tangherlini black holes. A possible extension of focus would be to consider effective actions involving yet higher $(R_{\mu\nu})^n$ powers of the Ricci tensor. One can directly see, however, that the Lichnerowicz eigenfunctions considered in this paper will continue to play similar dual r\^oles in such extended models. The fourth-order metric fluctuation equation \eqref{eq:metricfluct} arises from a second variation of the effective action: the first variation yields the field equation, while the second yields \eqref{eq:metricfluct}. The presence of $(R_{\mu\nu})^{n\ge3}$ terms in the effective action would not change anything in 
such a fluctuation analysis about an initial Ricci-flat background because 
a second variation would leave untouched $R_{\mu\nu}$ factors in the 
fluctuation equation, which would consequently all vanish.

\newpage

\section*{Acknowledgments}

We would like to thank Carl Bender, Jorge Santos, Maria Rodriguez and Toby Wiseman for helpful discussions. For warm hospitality during the course of the work, C.N.P.\ and K.S.S.\ would like to thank the Mitchell Family Foundation for the Brinsop Court workshop on strings and cosmology and the Galileo Galilei Institute; K.S.S. would also like to thank the Mitchell Institute of Texas A\&M University, the Isaac Newton Institute, the Albert Einstein Institute and Beijing Normal University. The work of H.L. was supported in part by NSFC grants 11175269, 11235003 and 11475024; the work of C.N.P. was supported in part by DOE grant DE-FG02-13ER42020;  the work of K.S.S. was supported in part by the STFC under Consolidated Grant ST/ST/L00044X/1 and the work of A.P. was supported by an STFC studentship.


\begin{thebibliography}{99}

\bibitem{Stelle:1976gc}
  K.S.~Stelle,
  ``Renormalization of higher derivative quantum gravity,''
  Phys.\ Rev.\ D {\bf 16} (1977) 953.
  doi:10.1103/PhysRevD.16.953.

\bibitem{Stelle:1977ry}
  K.S.~Stelle,
  ``Classical gravity with higher derivatives,''
  Gen.\ Rel.\ Grav.\  {\bf 9} (1978) 353.
  doi:10.1007/BF00760427.

\bibitem{Lu:2015cqa}
  H.~L\"u, A.~Perkins, C.N.~Pope and K.S.~Stelle,
  ``Black holes in higher-derivative gravity,''
  Phys.\ Rev.\ Lett.\ {\bf 114} (2015) 171601
  [arXiv:1502.01028 [hep-th]].

\bibitem{Lu:2015psa}
  H.~L\"u, A.~Perkins, C.N.~Pope and K.S.~Stelle,
  ``Spherically Symmetric Solutions in Higher-Derivative Gravity,''
  Phys.\ Rev.\ D {\bf 92} (2015) no.12,  124019
  doi:10.1103/Phys RevD.92.124019
  [arXiv:1508.00010 [hep-th]].

\bibitem{Perkins:2016imn}
  A.~Perkins,
  ``Static spherically symmetric solutions in higher derivative gravity,''
  https://spiral.imperial.ac.uk:8443/handle/10044/1/44072\,.

\bibitem{Goldstein:2017rxn}
  K.~Goldstein and J.J.~Mashiyane,
  ``Ineffective higher derivative black hole hair,''
  [arXiv:1703.02803 [hep-th]].

\bibitem{Gross:1982cv}
  D.J.~Gross, M.J.~Perry and L.G.~Yaffe,
  ``Instability of flat space at finite temperature,''
  Phys.\ Rev.\ D {\bf 25} (1982) 330.
  doi:10.1103/PhysRevD.25.330\,.

\bibitem{Whitt:1985ki}
  B.~Whitt,
  ``The stability of schwarzschild black holes in fourth order gravity,''
  Phys.\ Rev.\ D {\bf 32} (1985) 379.
  doi:10.1103/PhysRevD.32.379\'.

\bibitem{Gregory:1993vy}
  R.~Gregory and R.~Laflamme,
  ``Black strings and p-branes are unstable,''
  Phys.\ Rev.\ Lett.\  {\bf 70} (1993) 2837
  doi:10.1103/PhysRevLett.70.2837
  [hep-th/9301052].

\bibitem{Babichev:2013una}
  E.~Babichev and A.~Fabbri,
  ``Instability of black holes in massive gravity,''
  Class.\ Quant.\ Grav.\  {\bf 30} (2013) 152001
  doi:10.1088/0264-9381/30/15/152001
  [arXiv:1304.5992 [gr-qc]].

\bibitem{Brito:2013wya}
  R.~Brito, V.~Cardoso and P.~Pani,
  ``Massive spin-2 fields on black hole spacetimes: Instability of the Schwarzschild and Kerr solutions and bounds on the graviton mass,''
  Phys.\ Rev.\ D {\bf 88} (2013) no.2,  023514
  doi:10.1103/PhysRevD.88.023514
  [arXiv:1304.6725 [gr-qc]].

\bibitem{Myung:2013doa}
  Y.S.~Myung,
  ``Stability of Schwarzschild black holes in fourth-order gravity revisited,''
  Phys.\ Rev.\ D {\bf 88} (2013) no.2,  024039
  doi:10.1103/PhysRevD.88.024039
  [arXiv:1306.3725 [gr-qc]].

\bibitem{Zerilli:1970se}
  F.J.~Zerilli,
  ``Effective potential for even parity Regge-Wheeler gravitational perturbation equations,''
  Phys.\ Rev.\ Lett.\  {\bf 24} (1970) 737.
  doi:10.1103/PhysRevLett.24.737\,.

\bibitem{Nelson:2010ig}
  W.~Nelson,
  ``Static solutions for 4th order gravity,''
  Phys.\ Rev.\ D {\bf 82} (2010) 104026
  [arXiv:1010.3986 [gr-qc]].

\bibitem{Myung:2013cna}
  Y.S.~Myung,
  ``Unstable Schwarzschild-Tangherlini black holes in fourth-order gravity,''
  Phys.\ Rev.\ D {\bf 88} (2013) no.8,  084006
  doi:10.1103/PhysRevD.88.084006
  [arXiv:1308.3907 [gr-qc]].

\bibitem{Liu:2011kf}
  H.~Liu, H.~L\"u and M.~Luo,
{\it On black hole stability in critical gravities,}
  Int.\ J.\ Mod.\ Phys.\ D {\bf 21}, 1250020 (2012)
  doi:10.1142/S0218271812500204
  [arXiv:1104.2623 [hep-th]].

\bibitem{Fan:2014ala}
  Z.Y.~Fan and H.~L\"u,
{\it Thermodynamical first laws of black holes in quadratically-extended gravities,}
  Phys.\ Rev.\ D {\bf 91} (2015) no.6,  064009
  doi:10.1103/PhysRevD.91.064009
  [arXiv:1501.00006 [hep-th]].

\bibitem{Wald:1993nt}
  R.M.~Wald,
  ``Black hole entropy is the Noether charge,''
  Phys.\ Rev.\ D {\bf 48} (1993) 3427
  doi:10.1103/PhysRevD.48.R3427
  [gr-qc/9307038].

\bibitem{Iyer:1994ys}
  V.~Iyer and R.~M.~Wald,
  ``Some properties of Noether charge and a proposal for dynamical black hole entropy,''
  Phys.\ Rev.\ D {\bf 50} (1994) 846
  doi:10.1103/PhysRevD.50.846
  [gr-qc/9403028].

\bibitem{Schutz:1985zz}
  B.F.~Schutz and C.M.~Will,
  ``Black hole normal modes: a semianalytic approach,''
  Astrophys.\ J.\  {\bf 291} (1985) L33.
  doi:10.1086/184453\,.

\bibitem{Iyer:1986np}
  S.~Iyer and C.M.~Will,
  ``Black hole normal modes: a {WKB} approach. 1. foundations and application of a higher order {WKB} analysis of potential barrier scattering,''
  Phys.\ Rev.\ D {\bf 35} (1987) 3621.
  doi:10.1103/PhysRevD.35.3621\,.

\bibitem{BenderOrszag}
 C.M.~Bender and S.A.~Orszag, {\it Advanced Mathematical Methods for Scientists and Engineers} (McGraw-Hill, New York 1978; Springer 2010).

\bibitem{Gubser:2000ec}
  S.S.~Gubser and I.~Mitra,
  ``Instability of charged black holes in Anti-de Sitter space,''
  hep-th/0009126.

\bibitem{Gubser:2000mm}
  S.S.~Gubser and I.~Mitra,
  ``The evolution of unstable black holes in anti-de Sitter space,''
  JHEP {\bf 0108} (2001) 018
  doi:10.1088/1126-6708/2001/08/018
  [hep-th/0011127].

\bibitem{Reall:2001ag}
  H.S.~Reall,
  ``Classical and thermodynamic stability of black branes,''
  Phys.\ Rev.\ D {\bf 64} (2001) 044005
  doi:10.1103/PhysRevD.64.044005
  [hep-th/0104071].

\bibitem{Hollands:2012sf}
  S.~Hollands and R.M.~Wald,
  ``Stability of black holes and black branes,''
  Commun.\ Math.\ Phys.\  {\bf 321} (2013) 629
  doi:10.1007/s00220-012-1638-1
  [arXiv:1201.0463 [gr-qc]].

\bibitem{Wald:2014bia}
  R.M.~Wald,
  ``Dynamic and thermodynamic stability of black holes and black branes,''
  Fundam.\ Theor.\ Phys.\  {\bf 177} (2014) 229
  doi:10.1007/978-3-319-06349-2\_10\,.

\bibitem{Green:2015kur}
  S.R.~Green, S.~Hollands, A.~Ishibashi and R.M.~Wald,
  ``Superradiant instabilities of asymptotically anti-de Sitter black holes,''
  Class.\ Quant.\ Grav.\  {\bf 33} (2016) no.12,  125022
  doi:10.1088/0264-9381/33/12/125022
  [arXiv:1512.02644 [gr-qc]].

\end{thebibliography}
\end{document}